\documentclass[a4paper,12pt]{article}
\pdfoutput=1
\usepackage{jcappub}
\usepackage[demo,export]{adjustbox}
\usepackage{amsmath}
\usepackage{hhline}
\usepackage{bm}
\usepackage{xcolor,color,colortbl}
\usepackage[utf8]{inputenc}
\usepackage{hyperref}
\usepackage{url}
\usepackage{graphicx}
\usepackage{multirow,bigdelim}
\usepackage{amssymb}
\usepackage[amssymb]{SIunits}
\usepackage{bbold}
\usepackage{verbatim}
\usepackage{fontawesome5}
\usepackage{orcidlink}
\usepackage{float}

\usepackage{caption}
\usepackage{subcaption}
\usepackage{longtable}
\usepackage{booktabs}
\usepackage{makecell}
\usepackage{array}

\newcommand{\bs}[1]{\boldsymbol{#1}}

\compress

\title{Cosmological inference from a joint DESI DR1 full-shape power spectrum and bispectrum analysis}

\author[1]{Caroline Guandalin\,}
\author[2,3]{Prakhar Bansal\,}
\author[4]{Pedro Carrilho\,}
\author[5]{Alejandro Aviles\,}
\author[1]{Mike (Shengbo) Wang\,}
\author[1]{Marcos Pellejero-Ibañez\,}
\author[6]{Aaditya Sarma\,}
\author[7]{Jaide Swanson\,}
\author[8]{Marco Bonici\,}
\author[1]{Florian Beutler\,}
\author[9]{Arnaud de Mattia\,}
\author[7]{Hee-Jong Seo}

\affiliation[1]{Institute for Astronomy, University of Edinburgh, Royal Observatory, Blackford Hill, Edinburgh EH9 3HJ, UK}

\affiliation[2]{Leinweber Center for Theoretical Physics, University of Michigan, 450 Church Street, Ann Arbor, Michigan 48109-1040, USA}

\affiliation[3]{University of Michigan, 500 S. State Street, Ann Arbor, MI 48109, USA}

\affiliation[4]{Centre for Astrophysics Research, University of Hertfordshire, UK}

\affiliation[5]{Instituto de Ciencias F\'{\i}sicas, Universidad Nacional Aut\'onoma de M\'exico, Av. Universidad s/n, Cuernavaca, Morelos, C.~P.~62210, M\'exico}

\affiliation[6]{Astronomy Centre, University of Sussex, Falmer, Brighton BN1 9QH, UK}

\affiliation[7]{Department of Physics \& Astronomy, Ohio University, Athens, OH 45701, USA}

\affiliation[8]{Waterloo Centre for Astrophysics, University of Waterloo, Canada}

\affiliation[9]{IRFU, CEA, Universit\'{e} Paris-Saclay, F-91191 Gif-sur-Yvette, France}

\emailAdd{caroline.guandalin@roe.ac.uk}

\abstract{
The galaxy bispectrum directly probes the non-linear gravitational evolution of large-scale structure (LSS) and can break parameter degeneracies that remain in power-spectrum analyses. We present a joint full-shape cosmological analysis of three luminous red galaxy (LRG) redshift bins and the quasar (QSO) sample from the first Data Release (DR1) of the Dark Energy Spectroscopic Instrument (DESI). We model the redshift-space power spectrum at one loop and the tree-level bispectrum within the Effective Field Theory of LSS. The bispectrum is decomposed in the Tripolar Spherical Harmonics basis, for which the convolution with the survey window function can be formulated as a direct linear transformation of the theoretical multipoles. Our power-spectrum constraints are in good agreement with the official DESI DR1 full-modelling results. We investigate the impact of including the bispectrum in the inference, finding that the monopole substantially improves the constraints on the cold dark matter density and amplitude of matter fluctuations by 9-18\% and 8-20\%, respectively, in the individual-tracer analyses. The corresponding reductions are 15\% and 10\% for the combined LRG sample, and 6\% and 4\% when all tracers are combined. In a restricted test using the first LRG bin, the bispectrum quadrupole changes the marginalised uncertainties by only a few percent. Extending the analysis to $w_0w_a$CDM substantially broadens the cosmological posteriors, while the bispectrum produces only a mild change in the allowed dark-energy parameter region, which remains sensitive to the adopted prior ranges. Our results demonstrate the potential of higher-order clustering statistics to improve cosmological constraints, while providing a framework for incorporating the bispectrum into full-shape analyses of current and future spectroscopic galaxy surveys.
}

\begin{document}
\maketitle

\flushbottom

\section{Introduction} \label{sec:introduction}

Galaxy redshift surveys measure the three-dimensional distribution of biased tracers and provide one of the main probes of cosmic expansion and the growth of structure. While baryon acoustic oscillation (BAO) measurements provide a robust standard ruler, full-shape analyses use the scale dependence of the clustering power spectrum and its redshift-space anisotropies to extract additional information on the matter density, the amplitude of fluctuations, the growth rate, and galaxy bias \cite{desiBAO2025,DR1Pk}. Much of this additional information can be obtained from the scales where the density field is no longer in the linear regime, making an accurate description of non-linear structure formation essential.

Non-linear gravitational collapse of matter is responsible for the interaction of different scales, which in Fourier space is manifested through the coupling of different modes $\bs{k}_i$. At late times, these interactions generate non-negligible deviations from the nearly Gaussian statistics expected from the linear evolution of the initial conditions \cite{planck2018_PNG}. Quantifying the agreement of such gravitationally induced non-Gaussianities with theoretical predictions is essential for testing the assumptions and validity of our cosmological models.

We can study how these non-Gaussianities are generated analytically by describing the non-linear gravitational evolution through a convolution of $n$ Fourier modes of the initial (linear) density field $\delta^{(1)}$, weighted by a kernel $F_n$ that describes the mode coupling induced by gravity at order $n$ in standard perturbation theory (SPT) \cite{fry1984,goroff1986,bernardeau2002}.

With the two-point function in Fourier space (namely, the power spectrum),\footnote{Here, and in what follows, $\delta^{\rm D}(\bs{x})$ is the three-dimensional Dirac delta.}
\begin{equation}
    \langle \delta(\bs{k}_1) \delta(\bs{k}_2) \rangle = (2\pi)^3\delta^{\rm D}(\bs{k}_1+\bs{k}_2) P(\bs{k}_1), \label{eq:power_spec_def}
\end{equation}deviations from the linear prediction are already present due to loop corrections that come from the higher-order terms in the SPT expansion, such as $\delta^{(2)}$ and $\delta^{(3)}$. However, since a Gaussian random field is fully characterised by its two-point function, with all connected higher-order moments vanishing identically \cite{isserlis1918}, the three-point function, or bispectrum in Fourier space,
\begin{equation}
    \langle \delta(\bs{k}_1)\delta(\bs{k}_2) \delta(\bs{k}_3)\rangle = (2\pi)^3\delta^{\rm D}(\bs{k}_1+\bs{k}_2+\bs{k}_3)\, B(\bs{k}_1,\bs{k}_2,\bs{k}_3),\label{eq:bispec_def}
\end{equation}
is the lowest-order statistic sensitive to the non-Gaussian features, thus providing a direct probe of the non-linear gravitational interactions responsible for mode coupling in the late-time matter distribution. Therefore, in this work we consider a joint analysis of the galaxy bispectrum $B(\bs{k}_1,\bs{k}_2,\bs{k}_3)$ and power spectrum $P(\bs{k}_1)$. 

In addition to providing us with a better description of the distribution of galaxies, the bispectrum also helps us to constrain the galaxy bias expansion, which captures the galaxy formation process by a set of unknown bias parameters. Because of that, we are able to constrain non-linear bias contributions that lift degeneracies present in the linear theory, such as the linear bias and amplitude degeneracy in the power spectrum \cite{MVH,scoccimarro1999}, and high-order biases \cite{sefusatti2007,moradinezhad2021}. This is one of the main motivations for joint power-spectrum and bispectrum analyses in perturbative models of galaxy clustering \cite{gilmarin2015,eggemeier2021,philcox2022}.

One of the main goals of modern full-shape analyses of the power spectrum and bispectrum is to extend the modelling to the quasi-linear regime, while maintaining an accuracy compatible with the statistical precision of current galaxy redshift surveys. These scales contain a large fraction of the observable Fourier modes, which reduces the statistical uncertainties due to an improved sampling of the underlying density field. The primary advantage of current spectroscopic surveys, such as the Dark Energy Spectroscopic Instrument (DESI), is their large effective volume and broad redshift coverage, which reduce sample variance and provide several galaxy and quasar samples with complementary bias and redshift dependence. The number density of each sample then controls the balance between shot noise and sample variance on the scales included in the analysis \cite{desi2022overview,desiDR1,DR1Pk}.

However, the accurate modelling of the quasi-linear regime remains challenging. The loop corrections arising in the SPT receive contributions from small, non-linear scales where the perturbation expansion breaks down. The Effective Field Theory (EFT) of Large-Scale Structure (LSS) provides a physically motivated way of accounting for their impact on long-wavelength modes: unresolved short-distance dynamics is absorbed into counterterms and stochastic contributions consistent with the symmetries of the problem, making the perturbative description predictive on mildly non-linear scales \cite{baumann2012,carrasco2012,pajer2013,assassi2014,baldauf2015,Vlah:2015zda}.

Our primary goal in this work is to constrain the $\Lambda$CDM model and the galaxy bias parameters by conducting the first full-shape analysis of the joint galaxy power spectrum and bispectrum with DESI Year 1 data \cite{desiDR1}, accounting for the effects of survey geometry in a consistent way via the Tripolar Spherical Harmonics (TripoSH) decomposition of the redshift-space bispectrum, and within the framework of EFT. We also consider the extension to a time-dependent dark-energy equation of state described by the Chevallier-Polarski-Linder (CPL) parametrisation, $w(a)=w_0+w_a(1-a)$ \cite{chevallier2001,linder2002}.

We structure this manuscript as follows. In Section~\ref{sec:modelling}, we describe our modelling choice, which is based on the EFT of LSS, for both the power spectrum and bispectrum. We also discuss the TripoSH decomposition in Section~\ref{sssec:triposh}, and we give a general overview of the window function convolution in Section~\ref{ssec:window_convolution}, leaving the specific details of the DESI DR1 power spectrum window treatment to Section~\ref{ssec:powerspectrum}. We provide a full description of the bispectrum window function in Appendix~\ref{app:convolution}. Sections~\ref{sec:data} and \ref{sec:simulations} introduce the main data catalogues and simulations employed in our analysis. In Section~\ref{sec:methods}, we present the methodology chosen: the Bayesian inference formalism, the likelihood and priors assumed, and the sampler and emulators chosen. Finally, the results are presented in Section~\ref{sec:results}, and we conclude in Section~\ref{sec:conclusions}. The validity of the pipeline, tested against $N$-body simulations, is left for Appendix~\ref{app:validation}.

\section{Modelling}\label{sec:modelling}

The statistical power of galaxy surveys increases rapidly as smaller scales are included, since the number of independent Fourier modes grows approximately as $k^2\,{\rm d}k$. The challenge is that these same scales are progressively affected by non-linear gravitational evolution, galaxy bias, and redshift-space distortions. Standard perturbation theory is controlled only while the long-wavelength density fluctuations remain small, and it breaks down as the wavenumber approaches the non-linear scale $k_{\rm NL}$, defined schematically by $\delta(k_{\rm NL})\sim{\cal O}(1)$. A central goal of modern large-scale-structure modelling is therefore to extend the range of scales that can be used in a controlled way, without attempting to describe the fully non-linear short-distance dynamics explicitly.

The EFT of LSS provides a systematic framework for this purpose. In this context, EFT means a long-wavelength description obtained after coarse-graining over short-scale modes. The long-wavelength dark-matter fields are described as an effective fluid coupled to gravity, while the impact of unresolved short-distance physics is absorbed into counterterms and stochastic contributions that supplement the perturbative expansion with all counterterms allowed by the symmetries of the problem: equivalence principle, rotational invariance, and mass and momentum conservation  \citep{mcdonald2009,baumann2012,carrasco2012,pajer2013,baldauf2015,assassi2014,Vlah:2015zda}; see \cite{desjacques2018,ivanov2022} for reviews. This makes the perturbative description renormalised and predictive on mildly non-linear scales, with an accuracy controlled by ratios such as $k/k_{\rm NL}$ and by the perturbative order used in the calculation.

In this work, we use this framework for the redshift-space galaxy power spectrum at one loop and for the galaxy bispectrum at tree level. Our implementation follows the \texttt{FOLPS} perturbative solver \citep{Aviles:2021que,Noriega:2022nhf,bansal2026}, which provides fast loop calculations and can include kernels beyond the Einstein--de Sitter approximation when required. Infrared (IR) resummation is used to account for the large-scale bulk displacements that smooth the baryon acoustic oscillation feature \cite{Senatore:2014via,Vlah:2015sea,ivanov2018}.

For biased tracers, the small-scale physics of galaxy formation is absorbed into a set of bias coefficients multiplying all operators allowed by the symmetries \citep{mcdonald2006,mcdonald2009,assassi2014,desjacques2018,mergulhao2022}. At fixed time $\tau$, the galaxy overdensity in real space can be written as
\begin{equation}\label{eq:realsp_density_field}
    \delta_g(\bs{r},\tau) = \frac{n_g(\bs{r},\tau)}{\bar n_g(\tau)}-1 = \sum_{\cal O} b_{\cal O}(\tau)\,[{\cal O}](\bs{r},\tau) + \epsilon(\bs{r},\tau) + \sum_{\cal O} \epsilon_{\cal O}(\bs{r},\tau)\,[{\cal O}](\bs{r},\tau).
\end{equation}
Here $\bs{r}$ denotes the comoving real-space position, $\bar n_g$ is the mean galaxy number density, $b_{\cal O}$ are bias parameters, and $[{\cal O}]$ denotes renormalised composite operators built from second spatial derivatives of the gravitational and velocity potentials. The fields $\epsilon$ and $\epsilon_{\cal O}$ describe stochastic, \textit{noise} contributions sourced by short-scale physics and tracer discreteness.

At leading order in spatial derivatives, and for the power spectrum at one loop and the bispectrum at tree level, a convenient operator basis is
\begin{equation}
    {\cal O}\in \left\{ \delta,\delta^2,K^2,O_{\rm td}\right\},
\end{equation}
with corresponding bias parameters
\begin{equation}
    \left\{b_1,b_2,b_{K^2},b_{\rm td}\right\}.
\end{equation}
The tidal operator is defined by $K^2\equiv K_{ij}K^{ij}$, where
\begin{equation}
    K_{ij}=\left(\frac{\partial_i\partial_j}{\nabla^2}-\frac{1}{3}\delta_{ij}\right)\delta,
\end{equation}
while
\begin{align}
   O_{\rm td} &= \frac{8}{21} K_{ij} \left( \frac{\partial_i \partial_j}{\nabla^2} - \frac{1}{3}\delta_{ij}\right) \left( \delta^2 - \frac{3}{2} K_{kl} K_{kl} \right). 
\end{align}
Higher-derivative operators, such as $\nabla^2\delta$, and stochastic terms are included separately when required by the perturbative order and by the precision of the data.

Equation~\eqref{eq:realsp_density_field} is written in real-space coordinates. In redshift surveys, however, the observed radial position is shifted by the line-of-sight component of the peculiar velocity. In the plane-parallel approximation, the mapping from real-space position $\bs{r}$ to redshift-space position $\bs{s}$ is
\begin{equation}
    \bs{s} = \bs{r} + \hat{\bs{z}}\, \frac{\bs{v}(\bs{r})\cdot\hat{\bs{z}}}{\mathcal H},
\end{equation}
where ${\cal H}=aH$ is the conformal Hubble rate, $a$ is the scale factor, $\bs{v}$ is the peculiar velocity field, and $\hat{\bs{z}}$ denotes the fixed line-of-sight direction. Number conservation under this mapping determines the redshift-space density contrast. In Fourier space,
\begin{equation}
    \delta_g^{(s)}(\bs{k}) = \int {\rm d}^3 r\, {\rm e}^{-i\bs{k}\cdot\bs{r}} \left[ 1+\delta_g(\bs{r}) \right] \exp\left[-i k_z \frac{v_z(\bs{r})}{\mathcal H}\right] - (2\pi)^3\delta^{\rm D}(\bs{k}) ,
\end{equation}
where $v_z\equiv \bs{v}\cdot\hat{\bs{z}}$ and $k_z\equiv \bs{k}\cdot\hat{\bs{z}}$.

Expanding this expression perturbatively gives the standard redshift-space kernels $Z_n$ \citep{kaiser1987,scoccimarro1999,bernardeau2002},
\begin{eqnarray}
    \delta^{(s)}_g(\bs{k}) &=& Z_1(\bs{k})\delta^{(1)}(\bs{k}) + \int{\rm d}^3q_1\, {\rm d}^3q_2 \,\,\delta^{\rm D}(\bs{k}-\bs{q}_{12}) Z_2(\bs{q}_1,\bs{q}_2)\, \delta^{(1)}(\bs{q}_1)\,\delta^{(1)}(\bs{q}_2)\\
    &+& \int{\rm d}^3q_1\, {\rm d}^3q_2\, {\rm d}^3q_3 \,\,\delta^{\rm D}(\bs{k}-\bs{q}_{123}) Z_3(\bs{q}_1,\bs{q}_2,\bs{q}_3)\, \delta^{(1)}(\bs{q}_1)\,\delta^{(1)}(\bs{q}_2)\,\delta^{(1)}(\bs{q}_3),\nonumber
\end{eqnarray}
where $Z_1(\bs{k},\hat{\bs{n}})$, $Z_2(\bs{k}_1,\bs{k}_2,\hat{\bs{n}})$, and $Z_3(\bs{k}_1,\bs{k}_2,\bs{k}_3,\hat{\bs{n}})$ are the first-, second-, and third-order redshift-space kernels. At linear order, the mapping to redshift space gives the Kaiser formula,
\begin{equation}
    Z_1(\bs{k},\hat{\bs{n}})\equiv Z_1(\mu)=b_1+f\mu^2,    
\end{equation}
where $b_1$ is the linear galaxy bias, $f$ is the logarithmic growth rate, and $\mu=\hat{\bs{k}}\cdot\hat{\bs{n}}$ is the cosine of the angle between the Fourier mode and the line of sight \citep{kaiser1987}. This term describes the coherent large-scale velocity contribution to redshift-space clustering and sets the leading anisotropic dependence of the power spectrum. The higher-order kernels $Z_2$ and $Z_3$ generalise this result beyond linear theory: they include the mode coupling generated by non-linear density and velocity evolution, together with the bias operators required for galaxies. The tree-level bispectrum depends only on $Z_1$ and $Z_2$, whereas the one-loop redshift-space power spectrum also requires $Z_3$.

\subsection{Power spectrum}\label{ssec:pk}
We consider the power spectrum at 1-loop in perturbation theory \cite{bansal2026}:
\begin{equation}\label{eq:P_EFT}
    P_{\rm EFT}(k,\mu) = P_\text{tree}(k,\mu) + P_\text{1-loop}(k,\mu) + P_\text{c.t.}^{\rm LO}(k,\mu) + P_{\epsilon\epsilon}(k,\mu), 
\end{equation}where
\begin{equation}
    P_\text{tree}(k,\mu) = Z_1^2(\mu) P_\text{lin}(k)
\end{equation}is the standard redshift-space distortions term, 
\begin{equation}
P_\text{c.t.}^\text{LO}(k,\mu)= (c_0 + c_2 \mu^2 + c_4 \mu^4)\, k^2\, P_
\text{lin}(k) 
\end{equation}is the leading-order EFT counterterm, which emerges from the integration of modes above the EFT cut-off limit,
\begin{equation}
P_{\epsilon\epsilon}(k,\mu) = \frac{1}{\bar{n}}(\alpha_0+\alpha_2 k^2 \mu^2 )
\end{equation}
accounts for the stochastic nature of small-scale modes and of the initial conditions, and 
\begin{equation}
P_\text{1-loop}(k,\mu) = \sum_{m=0}^{4} \sum_{n=0}^{m} f^{m}\mu^{2n} I_{nm}(k)
\end{equation}
is the one-loop correction. Here, the dependence on the growth rate $f$ and on the line-of-sight angle $\mu$ has been made explicit. The functions $I_{mn}(k)$ are $k$-dependent functions, loop contributions, multiplying the corresponding powers of $f$ and $\mu^2$. They include, for example, terms associated with the usual $P_{22}$- and $P_{13}$-type loop integrals, together with the additional redshift-space and bias contributions that enter at the same perturbative order. We evaluate these loop functions using FFTLog methods \citep{talman1978,hamilton2000,simonovic2018}: the linear power spectrum is approximated by a sum of complex power laws, $P_{\rm lin}(k) \simeq \sum_a c_a k^{\nu_a}$, which allows the loop integrals to be recast as sums over power-law contributions. In this representation, the dependence on the input linear spectrum is encoded in the FFT coefficients $c_a$, while the remaining kernel-dependent part is stored in precomputed matrices ${\sf M}$. The evaluation of $I_{mn}(k)$ is then reduced to matrix contractions and sums over FFT modes. We refer the reader to Section 3 of \citep{Aviles:2020cax,Noriega:2022nhf} for further details in the \texttt{FOLPS} implementation.

We account for the effects of non-linear evolution of modes close to the baryon acoustic oscillations (BAO) scale $\ell_{\rm BAO}\sim104\,\,{\rm Mpc}/h$ via the IR-resummed power spectrum \cite{baldauf2015,ivanov2018, Noriega:2022nhf}:
\begin{equation}
    P^\text{IR-res}_{\rm EFT}(k,\mu) = {\cal D}(k,\mu)\, P_{\rm EFT}(k,\mu) + \left[1-{\cal D}(k,\mu)\right] P_{\rm EFT}^{\rm nw}(k,\mu) + k^2\,\Sigma^2_\text{tot}(k,\mu)\,{\cal D}(k,\mu) P_{\rm w}(k),
\end{equation}where ${\cal D}(k,\mu) \equiv {\rm e}^{-k^2 \Sigma^2_\text{tot}(k,\mu)}$, 
\begin{equation}
\Sigma^2_\text{tot}(k,\mu) = \big[1+f \mu^2 \big( 2 + f \big) \big]\Sigma^2 + f^2 \mu^2 (\mu^2-1)\, \delta\Sigma^2,  \end{equation}
\begin{equation}
\Sigma^2 = \frac{1}{6 \pi^2}\int_0^{k_\text{IR}} {\rm d}p \,P_{\rm nw}(p) \left[ 1 - j_0\left(p \,\ell_\text{BAO}\right) + 2 j_2 \left(p \,\ell_\text{BAO}\right)\right],
\end{equation}
\begin{equation}
\delta\Sigma^2 = \frac{1}{2 \pi^2}\int_0^{k_\text{IR}} {\rm d}p \,P_{\rm nw}(p)\,  j_2\left(p \,\ell_\text{BAO}\right),
\end{equation}$k_{\rm IR}\sim 0.4\,\,h/{\rm Mpc}$ is the scale separating the modes that should be resummed (responsible for the BAO broadening), from those that have no impact on the BAO feature, $j_0$ and $j_2$ are the spherical Bessel functions, $P_{\rm w}(k)$ is the BAO feature, such that $P_{\rm lin}(k)=P_{\rm nw}(k)+P_{\rm w}(k)$, and $P_{\rm EFT}^{\rm nw}(k,\mu)$ is obtained via Equation~\eqref{eq:P_EFT}, but using the smooth power spectrum $P_{\rm nw}$ instead of the linear one. 

\subsubsection{Legendre decomposition}\label{sssec:legendre}
In order to describe the anisotropies along the line of sight introduced by the redshift-space distortions, we decompose the galaxy power spectrum in redshift-space in terms of Legendre polynomials:
\begin{equation}\label{eq:legendre_expansion}
    P^{(s)}_g(k,\mu) = \sum_{\ell = 0}^{\infty} P_{\ell}^{(s)}(k)\mathcal{L}_{\ell}(\mu),
\end{equation}where
\begin{equation}
    P_{\ell}^{(s)}(k) = \frac{2\ell + 1}{2}\int_{-1}^1{\rm d}\mu\,\, P_g^{(s)}(k,\mu) \, {\cal L}_{\ell}(\mu)
\end{equation}are the coefficients of the expansion.

\subsection{Bispectrum}\label{ssec:bk}
We model the galaxy bispectrum as proposed in \cite{ivanov2022}:
\begin{eqnarray}\label{eq:bispectrum_model}
    B(\bs{k}_1,\bs{k}_2,\bs{k}_3) &=& \bigg[\left.2 Z_2(\bs{k}_1,\bs{k}_2) Z_1^\text{FoG}(\bs{k}_1)    Z_1^\text{FoG}(\bs{k}_2)  P^\text{IR-res}_{\rm lin}(\bs{k}_1) P^\text{IR-res}_{\rm lin}(\bs{k}_2)\right.\\
    &{}& \hspace{1cm} \left.+ B_\text{shot} \left(b_1 + 2\frac{P_\text{shot}}{B_\text{shot}}f\mu^2 \right) Z_1(\bs{k}_1)  P^\text{IR-res}_{\rm lin}(\bs{k}_1)+\text{ perms. }\right] + \tilde{P}_{\rm shot}^2 \nonumber
\end{eqnarray}
While this keeps the overall structure of the tree-level bispectrum, this model includes the leading-order effects of bulk flows through the IR-resummation $P^\text{IR-res}_{\rm lin}(k,\mu) \equiv P_{\rm nw}(k) + {\cal D}(k,\mu)\,P_{\rm w}(k)$, where $P_{\rm nw}$ is the smooth component of the linear matter power spectrum.~It also includes the Fingers-of-God (FoG) damping associated with the small-scale velocity dispersion via the kernel 
\begin{eqnarray}
    Z_1^{\rm FoG}(\bs{k}) = b_1 + f\mu^2 - c_1\mu^2\left(\frac{k}{k_{\rm IR}}\right)^2,
\end{eqnarray}where $c_1$ is a counterterm.

\pagebreak

\subsubsection{Tripolar spherical harmonics decomposition}\label{sssec:triposh}

Because the redshift-space bispectrum depends on the directions $\hat{\bs{k}}_1$, $\hat{\bs{k}}_2$, and $\hat{\bs{n}}$, we decompose $B(\bs{k}_1,\bs{k}_2,\hat{\bs{n}})$ in the tripolar spherical harmonics (``TripoSH'') basis, which form a complete orthonormal basis for functions depending on three directions \cite{varshalovich1988, sugiyama2019}:
\begin{equation}
\begin{aligned}
    B_{\ell_1\ell_2L}(k_1,k_2) &= N_{\ell_1\ell_2L}\,H_{\ell_1\ell_2L} \int\frac{{\rm d}\cos\omega\,{\rm d}\phi}{4\pi} \int \frac{{\rm d}\cos\theta_{12}}{2}\times\\
    &{}\hspace{10mm}\times \sum_M \begin{pmatrix}
\ell_1 & \ell_2 & L\\
0 & -M & M
\end{pmatrix} \,{\cal Y}_{\ell_2\,-M}^*(\theta_{12},0)\,{\cal Y}_{LM}^*(\omega,0)B(k_1,k_2,\theta_{12}, \omega,\phi),
\end{aligned}
\end{equation}where ${\cal Y}_{\ell m} \equiv \sqrt{4\pi/(2\ell+1)}\, Y_{\ell m}$ are the quasi-normalised spherical harmonics. In the following, we adopt the coordinate system in which $\bs{k}_1$ is aligned along the $z$-axis, and $\bs{k}_2$ lies in the $x$-$z$ plane:
\begin{align}
    \boldsymbol{k}_1 &= \left( 0,0,k_1 \right)\\
    \boldsymbol{k}_2 &= \left( -k_2\,\sqrt{1-\mu_2^2},0,k_2\,\mu_2\right)\\
    \boldsymbol{k}_3 &= \left( -k_1^x - k_2^x, -k_1^y-k_2^y,-k_1^z-k_2^z\right)\\
    \hat{\boldsymbol{n}} &= \left( \sqrt{1-\mu^2}\cos\phi,\sqrt{1-\mu^2}\sin\phi,\mu \right),
\end{align}where $\cos\theta_{12} \equiv \mu_2$ and $\cos\omega \equiv \mu$. Here, $B_{\ell_1\ell_2L}(k_1,k_2)$ are the bispectrum multipoles in the TripoSH basis, which are symmetric under the $(\ell_1,k_1) \leftrightarrow (\ell_2,k_2)$ exchange.

We opt for this decomposition primarily because of its properties under global rotations: by coupling the angular dependence $\ell_1$, $\ell_2$ and $L$ of $\hat{\bs{k}}_1$, $\hat{\bs{k}}_2$ and $\hat{\bs{n}}$, respectively, in states of total angular momentum $[(\ell_1\otimes\ell_2)_{\ell_{12}}\otimes L]_{L_{\rm tot}}$, a rotation operator becomes a block-diagonal Wigner $D$-matrix $D_{MM'}^{(L_{\rm tot})}$, acting only on the magnetic index $M$ \cite{varshalovich1988}. Rotational invariance implies that the bispectrum transforms as a scalar under simultaneous rotations of $\hat{\bs{k}}_1$, $\hat{\bs{k}}_2$, and $\hat{\bs{n}}$, so only the $L_{\rm tot} = 0$ contributes. Finally, parity symmetry $B(-\bs{k}_1,-\bs{k}_2,-\hat{\bs{n}}) = B(\bs{k}_1,\bs{k}_2,\hat{\bs{n}})$ is also straightforward to impose, since the basis has a defined parity of $(-1)^{\ell_1+\ell_2+L}$, such that the only contributions come from modes with even $\ell_1+\ell_2+L$.

We reinforce this choice, as discussed in \cite{wang2025} and summarised in Section~\ref{ssec:window_convolution}, because it leads to a particularly simple treatment with the window function, as opposed to a convolution in the so-called ``Scoccimarro'' basis \cite{scoccimarro2015}: in the latter, the bispectrum is projected into the Legendre multipoles $B_L(k_1,k_2,k_3)$, while the window convolution requires additional multipoles that must be summed over \cite[for example, see][]{pardede2022}. In contrast, the survey window function, the 3PCF and the bispectrum are all expanded in the same irreducible TripoSH basis. As a result, the window convolution only couples TripoSH multipoles among themselves, allowing the full operation to be expressed as a mixing matrix acting directly on the vector of coefficients $(\ell_1 \ell_2 L)$ \cite{sugiyama2019,wang2025}.

An additional advantage of working in this basis is related to the dimensionality reduction of the observed data vector \cite{wang2025}: the bispectrum multipoles in the TripoSH basis are obtained after integrating over the angular dependence of the triangle and the line-of-sight configuration, compressing the bispectrum into a two-dimensional statistic $B_{\ell_1\ell_2 L}(k_1, k_2)$. This decreases the size of the data vector and its covariance matrix, reducing the number of simulations required for the covariance matrix estimation.

\subsection{Alcock–Paczy\'{n}ski effect}

Finally, our modelling includes the so-called Alcock–Paczy\'{n}ski (AP) effect \cite{alcock1979}, which appears as geometric distortions in the mapping of observed redshift and angles into comoving distances under the assumption of a given fiducial cosmological model and parameters $\bs{\theta}_{\rm fid}$. The mismatch between the quantities computed at the fiducial values and the parameters being sampled $\bs{\theta}_{\rm test}$, can be described by two ``AP'' parameters, 
\begin{equation}
    q_\parallel = \frac{H(\bs{\theta}_{\rm fid},z)}{H(\bs{\theta}_{\rm test},z)}\frac{h_{\rm test}}{h_{\rm fid}}\,\text{ and }\, q_\perp = \frac{D_A(\bs{\theta}_{\rm test},z)}{D_A(\bs{\theta}_{\rm fid},z)}\frac{h_{\rm test}}{h_{\rm fid}}.
\end{equation}

The wave vector $\bs{k}_{\rm fid}$, which we take in units of ${\rm Mpc}/h$, is defined through the mapping from observables (redshift and angles) using the fiducial cosmological model. For the alternative parameter set (and/or model), the same observations can be mapped into a new coordinate system ``distorted'' as 
\begin{equation}
\bs{k}_{\rm test} = \frac{\bs{k}_\parallel^{\rm fid}}{q_\parallel} + \frac{\bs{k}_\perp^{\rm fid}}{q_\perp},
\end{equation}where $\bs{k}_\parallel^{\rm fid} = (\bs{k}^{\rm fid}\cdot\hat{\bs{n}})\,\hat{\bs{n}}$, and $\bs{k}_\perp^{\rm fid} = \bs{k}^{\rm fid}  - \bs{k}_\parallel^{\rm fid}$. When changing between cosmological parameters, the volume element changes as 
\begin{equation}
    \frac{{\rm d}^3k_{\rm test}}{{\rm d}^3k_{\rm fid}} = \frac{1}{q_\perp^2 q_\parallel},
\end{equation}such that the power spectrum multipoles get shifted by $(q_\perp^2 q_\parallel)^{-1}$, and the bispectrum multipoles by $(q_\perp^2 q_\parallel)^{-2}$.

\subsection{Window convolution}\label{ssec:window_convolution}
Consider the density field $\delta(\bs{x})$ defined in Equation~\eqref{eq:realsp_density_field}. We define the corresponding ``windowed'' field
\begin{equation}\label{eq:windowed_field}
    \hat{\delta}(\bs{x}) = W(\bs{x}) \,\delta(\bs{x})
\end{equation}through the so-called ``window function'' $W(\bs{x})$, which characterises the survey geometry and selection function. In configuration space, this map acts locally through pointwise multiplication. However, in Fourier space, this local operation induces a convolution, \begin{equation}
    \hat{\delta}(\bs{k}) = \int \frac{{\rm d}^3q}{(2\pi)^3} \,W(\bs{k}-\bs{q})\, \delta(\bs{q}),
\end{equation}which couples different Fourier modes and angular multipoles: if the window were isotropic, i.e., $W(\bs{q})=W(|\bs{q}|)$, the convolution would preserve angular momentum and each Legendre multipole would evolve independently. However, the anisotropic survey footprint breaks global rotational invariance, such that the convolution inevitably mixes different angular modes. 

The convolved power spectrum $\hat{P}_\ell(k)$ can be defined through the integral operator ${\cal W}_P$:
\begin{equation}\label{eq:convolved_pk}
    ({\cal W}_P P)_\ell(k)
    \equiv \sum_{\ell'}\int {\rm d}q \,\,W_{\ell\ell'}(k,q)P_{\ell'}(q),
\end{equation}where $W_{\ell\ell'}$ is the kernel that couples different multipoles due to anisotropies in the survey geometry. Similarly, the convolved bispectrum ${\hat B}_\lambda$, where $\lambda \equiv (\ell_1,\ell_2,L)$ in the TripoSH basis, can be defined via ${\cal W}_B$:
\begin{equation}\label{eq:convolved_bk}
    \left({\cal W}_B B\right)_\lambda(k_1,k_2)
    \equiv
    \sum_{\lambda'}
    \int {\rm d}q_1\,{\rm d}q_2\,\,
    W^B_{\lambda\lambda'}(k_1,k_2;q_1,q_2)\,
    B_{\lambda'}(q_1,q_2),
\end{equation}The operator kernel $W_{\lambda \lambda'}$ is similar to what the authors in \cite{pardede2022} would call a ``mixing matrix'' (see their Equation~3.8). 

The approach we take was developed in references \cite{sugiyama2019, wang2025}, which closely resembles the one adopted in power-spectrum studies \cite{wilson2017,beutler2017,beutler2021}. We can summarise the convolution operator ${\cal W}_B$ as the set of linear transformations
\begin{eqnarray}
    B\xrightarrow{{\cal T}_1}\zeta \xrightarrow{{\cal C}}\hat{\zeta}\xrightarrow{{\cal T}_2}\hat{B},
\end{eqnarray}where ${\cal T}_1$ is a  double spherical Bessel transform, ${\cal C}$ is the configuration-space window coupling, and ${\cal T}_2$ is a Fourier transform taking the windowed 3PCF to the convolved bispectrum.

Since each of these operators acts on a finite dimensional space (fixed by the multipoles and bins of the data vector), we can recast the whole transformation in terms of matrices ${\sf T}$ and ${\sf C}$, such that
\begin{eqnarray}
    \hat{\bs{B}}=({\sf T}_2\,\circ\,{\sf C}\,\circ\,{\sf T}_1)\,{\bs B} \equiv {\sf W}_B\, \bs{B},
\end{eqnarray}where the whole process has been defined in terms of a window matrix ${\sf W}:{\bs B}\rightarrow\hat{\bs B}$. This linear algebra formulation was also developed in \cite{pardede2022}, but their result is given in terms of a mixing matrix which is a continuous linear operator.

We describe the specific details for the convolution of the DESI DR1 power spectrum measurements in Section~\ref{ssec:powerspectrum}. Since the bispectrum was not an ``official'' by-product of DR1, we give a detailed description of the bispectrum convolution series in Appendix~\ref{app:convolution}.

\section{DESI DR1 data} \label{sec:data}
In this work, we use the public DESI Data Release 1 (DR1) dataset \cite{desiDR1}, resulting from the DESI observations conducted between May 14 (2021) and June 14 (2022) on the Mayall Telescope at Kitt Peak, Arizona. We focus on the targets detected by the so-called ``dark-time programme'', namely the luminous red galaxies (LRGs, \cite{LRG}) and quasars (QSOs, \cite{QSO}).\footnote{In our analysis, we neglect the emission line galaxies (ELGs)  for simplicity.}

\begin{table}
\centering
\caption{Samples considered in this work. In our likelihood analysis, we fix the comoving number density of galaxies and quasars to $\bar{n}(z) = 10^{-4}\,\,(h/{\rm Mpc})^3$ .}
\label{tab:data}
\renewcommand{\arraystretch}{1.2}
\begin{tabular}{lccc}
{}\\
\toprule
\textbf{Sample} & \textbf{Redshift} & $\bs{N_g}$ & $\bs{z_{\rm eff}}$ \\
\midrule
LRG1 & $0.4 < z < 0.6$ & 506\,905 & 0.510 \\
\midrule
LRG2 & $0.6 < z < 0.8$ & 771\,875 & 0.706 \\
\midrule
LRG3 & $0.8 < z < 1.1$ & 859\,824 & 0.919 \\
\midrule
QSO & $ 0.8 < z < 2.1$  & 856\,652 & 1.491 \\
\bottomrule
\end{tabular}
\end{table}

Three-dimensional redshift catalogues are constructed from the targets' angular positions (RA, Dec) and the spectroscopic redshifts $z$ measured with DESI's pipeline \cite{zpipe}. For clustering analyses, DESI provides random Poissonian catalogues $n_{\rm r}(\bs{x})\sim{\rm Pois}[\nu = \bar{n}(z)]$ that match the survey angular mask (uniform distribution in the angular coordinates) and the empirical radial selection function $\bar{n}(z)$ of each tracer sample. They are generated with $\alpha^{-1} = N_{\rm r}/N_{\rm g} \gtrsim 100$, that is, $\gtrsim 100$ times denser than the data catalogue in order to minimise shot-noise contributions from the random field.\footnote{Here, $N_r$ and $N_g$ are the total number of random and data points, respectively.}

The random catalogues are used to estimate the survey selection function, and they isolate the intrinsic galaxy fluctuations in the weighted overdensity field that enters the power spectrum and bispectrum estimators,
\begin{equation}
    F(\bs{x}) = w(\bs{x})[n_{\rm g}(\bs{x}) - \alpha\, n_{\rm r}(\bs{x})]\,,
\end{equation}where $n_{\rm g}(\bs{x})$ and $n_{\rm r}(\bs{x})$ denote the data and random number-density fields, respectively, and $w(\bs{x}) = w_{\rm tot}(\bs{x}) \,w_{\rm FKP}(\bs{x})$ \cite{adame2025} is the total clustering weight applied to the catalogues, $w_{\rm tot}$ accounts for observational systematics,
fibre-assignment completeness, and redshift failures, and $w_{\rm FKP}$ is the FKP weight which optimises the balance between shot noise and sample variance for a specific scale in the power-spectrum estimation~\cite{FKP}.

The random catalogues are also used to determine the window function of the survey required to convolve theoretical predictions before they are compared with the observations (see section~\ref{ssec:window_convolution}). For further details on the specific data collection methodology employed by DESI, we refer the interested reader to \cite{DR1Pk} and references therein. 

For our main analysis, we consider the monopole and quadrupole of the power spectrum, and the monopole of the bispectrum, in the basis described in Sections~\ref{sssec:legendre} and \ref{sssec:triposh}, respectively. As we discuss in Section~\ref{ssec:bispectrum-analysis}, the bispectrum quadrupole does not add significant statistical gains and is therefore omitted from the analysis.

\begin{figure}[ht]
    \centering
    \includegraphics[width=\linewidth]{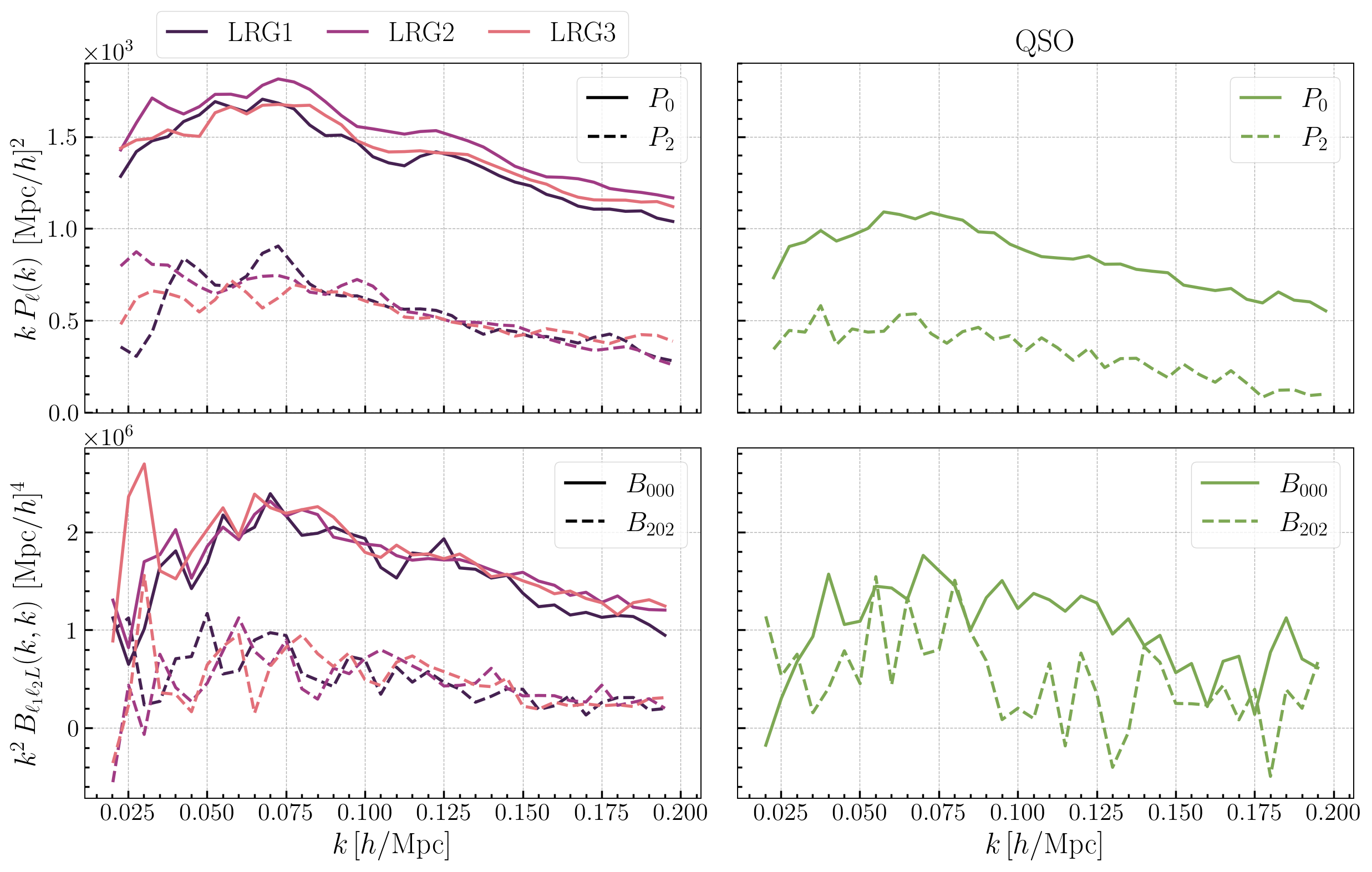}
    \caption{Power spectrum (`$\theta$-cut+rotated+RIC+AIC') and bispectrum measurements used in our full-shape analysis of the DESI DR1 LRG and QSO samples. The upper panels show the power spectrum monopole $P_0$ (solid lines) and quadrupole $P_2$ (dashed lines), while the lower panels show the diagonal bispectrum multipoles $B_{000}$ (solid lines) and $B_{202}$ (dashed lines). Left panels correspond to the three LRG redshift bins, $0.4<z<0.6$ (LRG1), $0.6<z<0.8$ (LRG2), and $0.8<z<1.1$ (LRG3), shown in different colours, and the right panels show the QSO sample in the range $0.8<z<2.1$. Notice that these are the ``GC-combined'' statistics (Eq.~\ref{eq:summary_GCcomb}).}
    \label{fig:DR1_LRG_QSO}
\end{figure}

As described in \cite{adame2025}, the measurements are performed in the South (SGC) and North Galactic Cap (NGC). We employ the same convention as in \cite{DR1Pk} and combine the power spectrum and bispectrum measurements into a weighted average:
\begin{equation}\label{eq:summary_GCcomb}
    S_a^\text{GC-comb}=\frac{\sum_{\rm GC} I_p^{\rm GC} S_a^{\rm GC}}{\sum_{\rm GC} I_p^{\rm GC}},
\end{equation}where ${\rm GC} \in \{\mathrm{SGC},\mathrm{NGC}\}$, and $S_a=P_\ell(k)$ or $S_a=B_{\ell_1\ell_2L}(k_1,k_2)$. The normalisation factors are given by 
\begin{equation}
    I_p = \alpha \sum_{i\in R} n_{z,i}^{p-1}\, w_{{\rm tot},i}\, w_{{\rm FKP},i}^{p}, 
\end{equation}with $n_{z,i}$ being the unweighted expected number density at the particle position, $p=2$ for the power spectrum, and $p=3$ for the bispectrum.

Due to the linear property of the window matrices, for both the power spectrum and bispectrum, as described in Section \ref{ssec:window_convolution}, the same weighted average is performed on the final window matrices, such that we compute a single theory vector which is then automatically converted into the GC-combined theory via the window convolution. We combine the windows in the way described by Equation~\eqref{eq:summary_GCcomb}, with $S_a\in \{{\sf W}_P,{\sf W}_B\}$.

\subsection{Power spectrum measurements}\label{ssec:powerspectrum}

We use the official DESI DR1 `$\theta$-cut+rotated+RIC+AIC' power spectrum measurements (see top panels of Figure~\ref{fig:DR1_LRG_QSO}). These are obtained from a $\theta$-cut estimator, $\hat{P}_\ell^{\rm cut} \equiv \hat{P}_\ell-\Delta P_\ell^{\theta_{\rm cut}}$, that removes the contributions $\Delta P_\ell^{\theta_{\rm cut}}$ to the power spectrum coming from pairs whose angular separation is below the fibre-collision scale $\theta<\theta_{\rm cut} = 0.05$ deg, offering an alternative to a forward modelling approach that would instead try to model the DESI fibre assignment~\cite{pinon2025}. These measurements also include corrections for the radial and angular integral constraints (RIC and AIC). For more details, see \cite{adame2025}.

The restriction of the angular domain imposed by the $\theta$-cut estimator directly impacts the window function. In configuration space, $W_{\rm cut} \sim W_{\rm raw}\,\Theta(\theta - \theta_{\rm cut})$, where $\Theta(x)$ is the Heaviside step function. In Fourier space, this angular cut will lead to sinc-like oscillations that decay slowly and thus pick up high-$k_{\rm theory}$ modes. That is, without the $\theta$-cut, the window function would be compact in the theory modes, such that a truncation at $k_{\rm max}$ could be safely done. 

To mitigate this effect, a ``compactification'' in $k$-space is applied to the power spectrum window matrix, such that most of its contributions are obtained from $k_{\rm obs} \approx k_{\rm theory}$. For that, the authors of \cite{pinon2025} considered the invertible linear transformation ${\sf M}:\mathbb{R}^{M_{\rm obs}}\rightarrow\mathbb{R}^{M_{\rm obs}}$, which gives combinations of the observed multipoles that are less sensitive to high-$k_{\rm theory}$ modes.\footnote{They called this transformation ${\sf M}$ a rotation, hence the name of the official DR1 power spectrum measurements.} Thus, ${\sf M}$ is such that 
\begin{equation*}
    {\sf W} \rightarrow \tilde{{\sf W}} = {\sf M}\,{\sf W}
\end{equation*}is nearly diagonal. Because this transformation acts on the data space, as $\bs{d}\rightarrow\tilde{\bs{d}}={\sf M}\,\bs{d},\,\, {\sf W} \rightarrow \tilde{{\sf W}} = {\sf M}\,{\sf W}, \text{ and } \tilde{\sf C} = ({\sf M}\,{\sf C}\,{\sf M}^{\sf T})$, it is important to require invariance of the likelihood. It is straightforward to check that this is achieved if ${\sf M}$ has an inverse. 

Since the compactified window matrix $\tilde{\sf W}$ is obtained by truncating the response to high-$k_{\rm theory}$ modes, the convolved prediction will be affected by a residual uncertainty associated with the missing part of the theory vector. In \cite{pinon2025}, this residual contribution is modelled as a correction to the window matrix, ${\sf W}\rightarrow {\sf W}-\bs{m}_{\rm o} \bs{m}_{\rm t}^{\sf T}$, where $\bs{m}_{\rm t}$ describes the characteristic pattern of the missing high-$k_{\rm theory}$ modes and $\bs{m}_{\rm o}$ specifies how this missing signal projects onto the observed data vector. Marginalising over an unknown amplitude $s$ of this residual contribution is equivalent to treating it as an additional source of uncertainty, since the amplitude of this contribution, which depends on the part of the theory vector beyond the adopted $k_{\rm max}$, cannot be predicted once those modes are truncated. This induces an extra covariance term, ${\sf C}_s \equiv s^2 \bf{m}_{\rm o} \bf{m}_{\rm o}^{\sf T}$, which is accounted for in the final covariance matrix.

\begin{figure}[ht]
\begin{subfigure}{\textwidth}
    \centering
    \includegraphics[width=\textwidth]{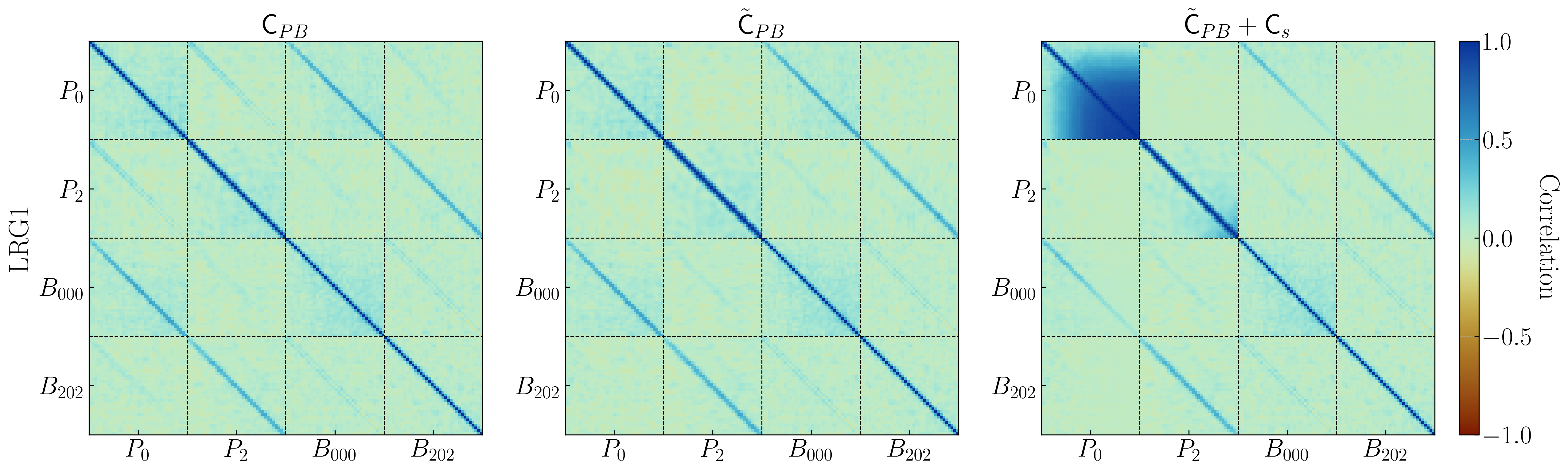}
    \label{fig:correlation_LRG1}
\end{subfigure}
\begin{subfigure}{\textwidth}
    \centering
    \includegraphics[width=\textwidth]{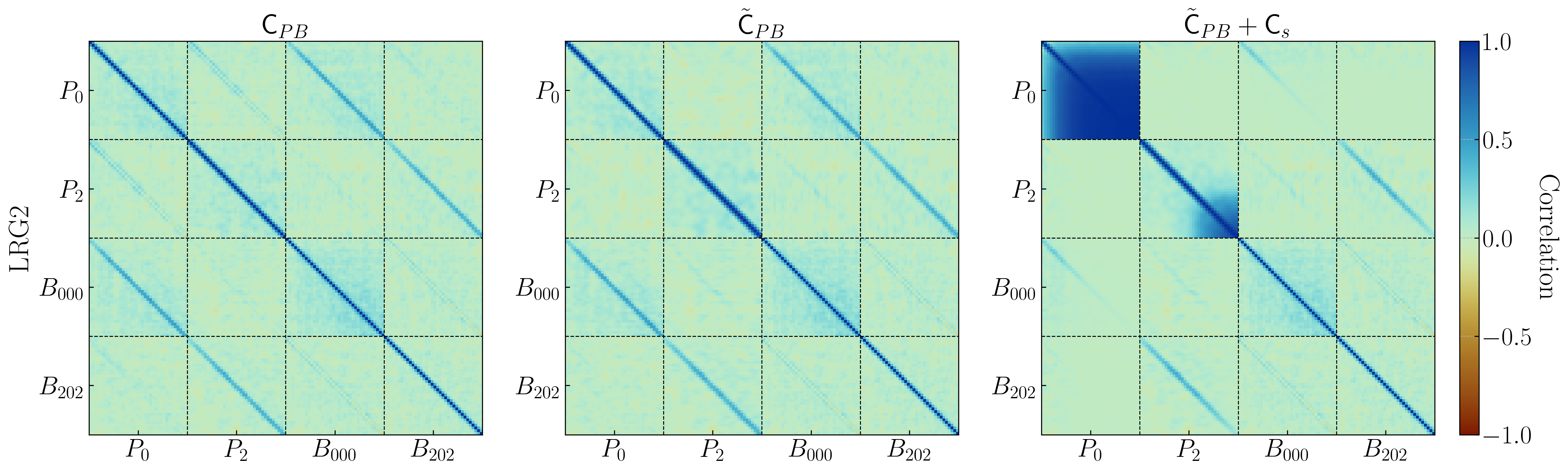}
    \label{fig:correlation_LRG2}
\end{subfigure}
\caption{Example of correlation matrices obtained from the LRG1 and LRG2 covariances, for the monopole and quadrupole of both power spectrum and bispectrum, within the range of scales considered in this work ($0.02<k<0.2\,\,h/{\rm Mpc}$). ${\sf C}_{PB}$ are the raw covariances obtained from the mock measurements (as described in Section \ref{ssec:cov}), $\tilde{\sf C}_{PB} = {\sf M}\,{\sf C}_{PB}\,{\sf M}^{\sf T}$ are the covariances with the linear transformation applied, and ${\sf C}_s = \sum_i s_i^2 {\bf m}_{o,i}{\bf m}_{o,i}^{\sf T}$ corresponds to the contributions coming from the marginalisation over the unknown amplitude $s$ used for the high-$k_{\rm theory}$ modes removal. }
\label{fig:covariances}
\end{figure}

This contribution, however, should not be interpreted as an additional independent noise in each Fourier mode: the same nuisance amplitude $s$ shifts all bins simultaneously according to $\bf{m}_{\rm o}$. As a result, the covariance between two observed bins $i$ and $j$ receives a contribution $(C_s)_{ij} = s^2 (m_{\rm o})_i(m_{\rm o})_j$. Since the template is largest at the high-$k$ modes affected by the window truncation, these modes acquire strong correlations after the marginalisation over $s$. Notice that in the $P$+$B$ covariance, this additional term modifies only the power-spectrum block: the $P$-$B$ cross-covariance is left unchanged. 

In Figure~\ref{fig:covariances}, we show the successive impact of these operations on the joint power-spectrum and bispectrum covariances for the LRG1 and LRG2 bins. The left panels show the correlation measured directly from the covariance of the mocks described in Section~\ref{ssec:cov}, while the middle panels show the effect of the linear transformation matrix ${\sf M}$, which mainly redistributes the correlations of the power-spectrum modes. Its effect on the bispectrum is given by the identity matrix $\mathbb{1} \equiv {\rm diag}(1,1,\ldots,1)$. The right panels additionally include the covariance contribution ${\sf C}_s$ associated with the marginalisation over the $s\,\bf{m}_{\rm o}$ template. The $P$-$B$ covariances are left unchanged by this additional term: the corresponding correlation coefficients are reduced because the power-spectrum variances entering their normalisation are increased by ${\sf C}_s$.

\subsection{Bispectrum measurements}
Our bispectrum measurements are obtained with the \texttt{Triumvirate} package \cite{sugiyama2018,sugiyama2019,wang2023,Wang:2023b}, using its FFT-based estimator for the TripoSH multipoles (see bottom panels of Figure~\ref{fig:DR1_LRG_QSO}).\footnote{\href{https://triumvirate.readthedocs.io/}{https://triumvirate.readthedocs.io/}} Particles are painted onto the FFT mesh using the Piecewise Cubic Spline (PCS) mass-assignment scheme (MAS).\footnote{The FFT mesh had the same dimensions $896\times1792\times896$ for all of our tracers; for this reason, the cell size is a tracer- and region-dependent quantity, varying from $\approx 1.7$ to $\approx 5.8\,\,{\rm Mpc}/h$.} The aliasing introduced by the MAS is treated in \texttt{Triumvirate} by compensating the Fourier-space field with the PCS window \cite{hockney2021}  
\begin{equation}
    W_{\rm PCS}(\bs{k}) = \left[{\rm sinc}\left(\frac{k_x\,\Delta}{2}\right){\rm sinc}\left(\frac{k_y\,\Delta}{2}\right){\rm sinc}\left(\frac{k_z\,\Delta}{2}\right)\right]^4,
\end{equation}where $\Delta$ is the grid spacing, and by applying the TSC aliasing correction to the shot-noise:
\begin{eqnarray}
    C_{\rm PCS}(\bs{k})  = 
    \prod_{i=x,y,z}\left[1 - \frac{4}{3}\sin^2\left(\frac{k_i\,\Delta}{2}\right) + \frac{2}{5} \sin^4\left(\frac{k_i\,\Delta}{2}\right)-\frac{4}{315}\sin^6\left(\frac{k_i\,\Delta}{2}\right)\right].
\end{eqnarray}

For each tracer, the catalogue is restricted to the chosen redshift interval (see Table~\ref{tab:data}) and converted from (RA, DEC, $z$) to Cartesian coordinates using the DESI fiducial cosmology (see Section \ref{ssec:nbody}). The bispectrum is measured, for each galactic cap, over the diagonal configurations $k_1 = k_2 = k$ in the range $0.005 < k < 0.3\,\,h/{\rm Mpc}$, with 60 equally-spaced bins with width $\Delta k=0.005\,\,h/{\rm Mpc}$. Throughout this work, we restrict the analysis to the lowest-order even-parity multipoles, namely $B_{000}$ and $B_{202}$, as motivated by \cite{sugiyama2019}, who have shown that $B_{202}$ has the largest quadrupole amplitude compared to $B_{112}$; however, its signal-to-noise is much lower than $B_{000}$. As we demonstrate in Section~\ref{ssec:bispectrum-analysis}, the quadrupole did not improve the cosmological constraints for our analysis set-up, when considering the LRG1 bin and the scale cut $k_{\rm max} = 0.08\,\,h/{\rm Mpc}$.

Unlike the power-spectrum measurements, our estimators do not employ the $\theta$-cut procedure described in \cite{pinon2025}. As we show in Appendix~\ref{app:convolution}, the residual differences between the window-convolved predictions and the mock measurements used to assess the accuracy of our approach are small over the scales used in our analysis. Therefore, we do not introduce an additional $\theta$-cut-like approach to the bispectrum measurements.\footnote{Readers interested in a $\theta$-cut approach to the bispectrum are invited to see \cite{chudaykin2026}.} Finally, the integral constraints are included in the bispectrum window matrices, if they have a non-negligible impact. For our samples, this was only the case for the quasars quadrupole (see Appendix~\ref{app:convolution}). 

\section{Simulations}\label{sec:simulations}

\subsection{\textit{N}-body simulations}\label{ssec:nbody}
We consider the \textsc{AbacusSummit} simulation suite \cite{abacus2021} and adopt the baseline $\Lambda$CDM cosmology as the reference case for consistency checks. The parameters are set to the mean of the full temperature, polarisation, and lensing (TT, TE, EE + lowE + lensing) data obtained from the Planck 2018 \texttt{Plik} results \cite{planck2018}: $\omega_{\rm b} = 0.02237$, $\omega_{\rm cdm} = 0.12$, $h = 0.6736$, $A_s = 2.0830 \times 10^{-9}$, $n_s = 0.9649$, $N_{\rm ur} = 2.0328$, $N_{\rm ncdm} = 1$, $\omega_{\rm ncdm} = 6.442 \times 10^{-4}$.

We use 25 independent realisations of this baseline cosmology generated in boxes of side $L = 2\,{\rm Gpc}/h$, with a particle mass of $2.1\times10^9\,M_\odot/h$. These simulations were built to generate mock catalogues reproducing the clustering of many DESI tracers, but we will only consider LRGs and QSOs: for the former, 50 snapshots have been generated (25 at $z=0.5$, 25 at $z=0.8$), while for the latter, we have 25 snapshots at $z=1.4$. 

From these snapshots, cubic mock catalogues were created to replicate the clustering of LRGs at redshifts $z=0.5$ (LRG1), $z=0.8$ (LRG2) and $z=1.1$ (LRG3) and of QSOs at $z=1.4$. These are part of the so-called ``Second Generation'' of mocks, which were created by populating the dark matter halos with the halo occupation distribution (HOD) model of \cite{zheng2007}, fitted to the final DESI early data release \cite{EDR}. The final periodic mocks reproduced the clustering of LRGs at $ 0.4 < z < 0.6$ (LRG1), $0.6 < z < 0.8$ (LRG2), and $0.8 < z < 1.1$ (LRG3), and quasars (QSO) between $0.8 < z < 2.1$. When constructing the redshift-space catalogues, the HOD accounted for an incompleteness factor $0 < f_{\rm ic} \leq 1$, and velocity-bias parameters for central and satellite galaxies \cite{yuan2024}. The number densities, in units of $h^3/{\rm Mpc}^3$, are $\bar{n}_{\rm LRG1} = 0.00070$, $\bar{n}_{\rm LRG2} = 0.00085$, $\bar{n}_{\rm LRG3} = 0.00081$, and $\bar{n}_{\rm QSO} = 0.00017$. In what follows, we refer to these cubic mocks simply as ``boxes''. 

Finally, two other sets of mocks were created, in order to reproduce the actual DESI observations: the ``complete'' and ``alternate Merged Target Ledgers'' (altMTL) mocks. From boxes described above, the complete catalogues were created by converting the Cartesian coordinates into the corresponding right ascension, declination and ``observed'' redshifts (we only consider those in redshift space), followed by the DESI footprint ``cut'': the angular positions were selected according to the area of observations, and the radial selection was applied in order to match the observed number density $n(z)$ of tracers. Because of that, we stress that these mocks are not proper light cones, in the sense that they do not carry any redshift evolution of the tracers. Finally, by applying fibre assignment of DESI observations to the complete mocks, the altMTL catalogues were generated with the same incompleteness as the real data, due to the fibre collisions. 

\begin{figure}[ht]
    \centering
    \includegraphics[width=\linewidth]{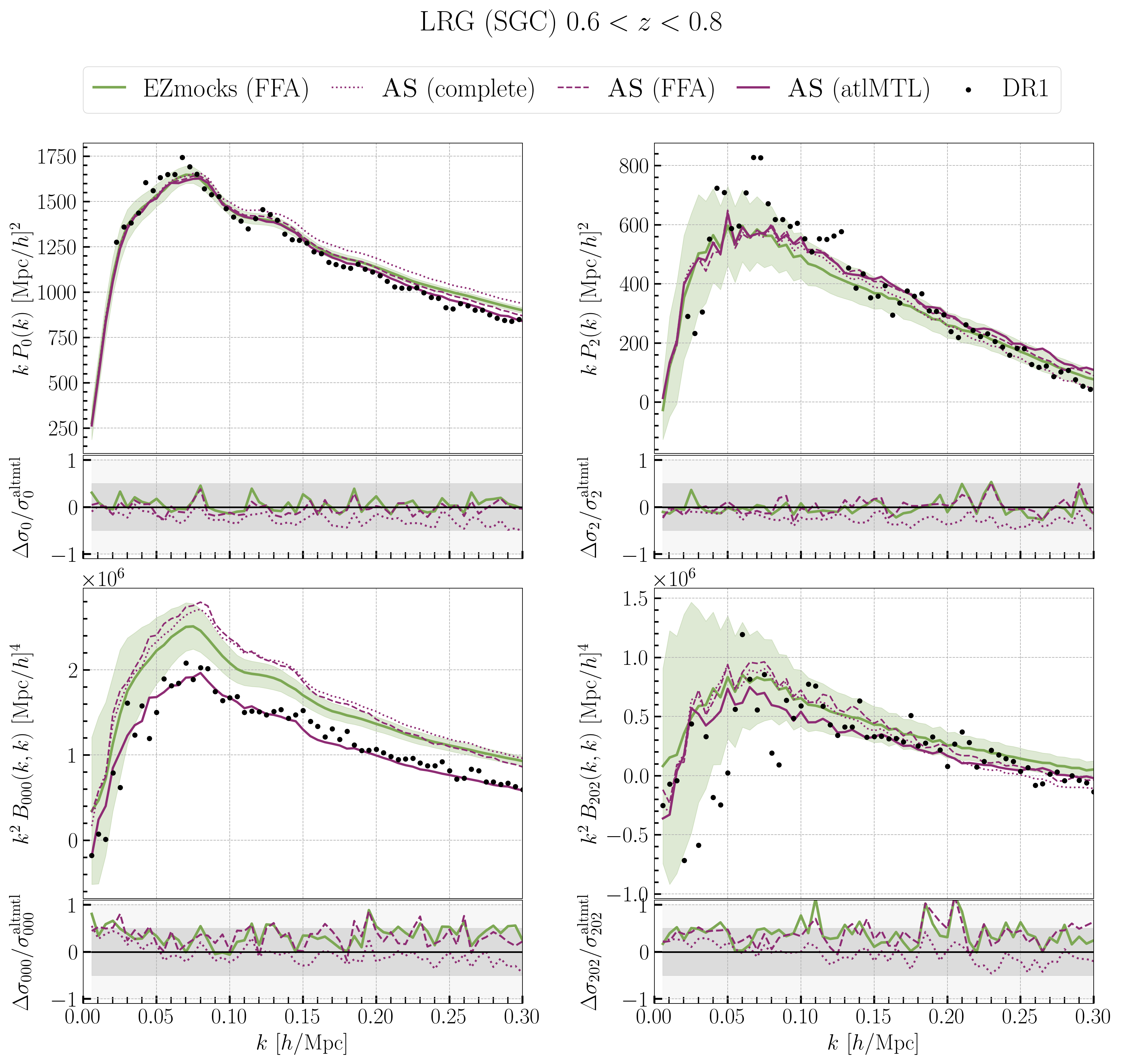}
    \caption{Comparison between different mocks for the LRG sample between $0.6 < z < 0.8$, for the SGC region. Green solid lines show the mean of 1000 EZmocks for the cut-sky case, with the fast-fibre assignment (FFA) applied, and the light-green shaded region shows their 1-$\sigma$ standard deviation. Magenta lines show the mean of 25 \textsc{AbacusSummit} cut-sky mocks for the ``complete'' (dotted) sample (i.e., no fibre assignment), and the ``FFA'' (dashed) and ``altMTL'' (solid) assignments. We compare these mocks with the DESI DR1 data (black points) which we fit in this work. Top panels: Monopole $P_0(k)$ (left) and quadrupole $P_2(k)$ (right) of the power spectrum. Bottom panels: Diagonal bispectrum monopole $B_{000}(k,k)$  (left) and quadrupole $B_{202}(k,k)$ (right). In the subpanels, the relative differences with respect to the AbacusSummit altMTL reference standard deviation are shown: $\Delta X/X_{\rm altMTL} = (X - X_{\rm altMTL})/X_{\rm altMTL}$, where $X=\lbrace \sigma_0, \sigma_2,\sigma_{000},\sigma_{202} \rbrace$. Shaded bands indicate the 0.5-$\sigma$ (dark) and 1-$\sigma$ (light) differences.}
    \label{fig:fibre}
\end{figure}

Figure~\ref{fig:fibre} shows an example of the power spectrum and bispectrum multipoles of these different sets of mocks for the second LRG bin ($0.6 < z < 0.8)$ in the SGC region. From these mock measurements, we can assess the impact of fibre collisions on the different statistics: it is most evident in the small-scale amplitude of the power spectrum monopole and the bispectrum. The ``complete’’ case (dotted), which does not include any fibre assignment, has a systematically higher amplitude of small-scale fluctuations because pairs with angular separation smaller than $\theta_{\rm cut} = 0.05\,{\rm deg}$ are not observed due to fibre collisions. As mentioned in Section~\ref{ssec:powerspectrum}, a $\theta$-cut window is applied to the power-spectrum theory vector in order to account for the effect of fibre collisions.

\subsection{Covariance}\label{ssec:cov}
The covariance matrix ${\sf C}$ is estimated from 1000 realisations of a periodic box of side $L_{\rm box} = 6\,\,{\rm Gpc}/h$, obtained via the \texttt{EZmocks} code \citep{chuang2014}. This code employs the Zel'dovich approximation to generate the dark matter density field, with an effective bias to describe the stochastic, non-local and non-linear contributions. These are calibrated to the \textsc{AbacusSummit} mocks to describe the galaxy field of interest.

In order to obtain realistic covariance matrices, the EZmocks were post-processed with an emulator that mimics the effect of fibre collisions: because applying the DESI fibre assignment algorithm to the 1000 mocks for each tracer and redshift bin was very time consuming, there are no ``altMTL'' catalogues for the EZmocks used to construct the covariance matrices. Instead, the incomplete set of EZmocks has been generated with the ``fast-fibre assignment'' (FFA) emulator described in \cite{bianchi2025}.

As shown in Figure~\ref{fig:fibre}, the FFA algorithm does not capture the effects of the realistic altMTL fibre assignment. However, the variances are all consistent if one accounts for fibre assignment. Therefore, for the purposes of our analysis, we use the covariances obtained from the 1000 FFA EZmocks. 

\section{Methodology} \label{sec:methods}

\subsection{Bayesian inference}\label{ssec:bayes}

In this work, we adopt the Bayesian framework to infer a set of $N_p$ parameters $\bs{\theta} = (\theta_1,\ldots,\theta_p)$ from a data vector $\bs{d} \in \mathbb{R}^{N_d}$ containing $N_d$ observations of the clustering statistics. We will decompose $\bs{\theta}$ into a set of parameters of interest $\bs{\theta}_*$ (e.g., the cosmological ones) and a set of nuisance parameters $\bs{\psi}$ (such as EFT counterterms and higher-order biases). Within this framework, $\bs{\theta}$ is treated as a random variable, such that we can assign it a probability density function $p(\bs{\theta}|\bs{D},M)$, called the posterior distribution, in contrast with the frequentist approach.\footnote{For a discussion on frequentist parameter inference in cosmology see, for example, \cite{holm2023, herold2025} and references therein.} We stress the fact that the distributions $p(\cdot\,|\,\cdot,M)$ are always conditioned by the model $M$ being tested. 

From Bayes' theorem, 
\begin{equation}
    p(\bs{\theta}|\bs{D},M) = \frac{p(\bs{D}|\bs{\theta},M)\,p(\bs{\theta},M)}{p(\bs{D},M)},
\end{equation}the posterior distribution $p(\bs{\theta}|\bs{d},M)$ for the observed realisation $\bs{d}$ of the random data vector $\bs{D}$ can then be estimated via some sampling algorithm.\footnote{We distinguish between the random data vector $\bs{D}$, representing an ensemble of possible observations generated by some stochastic process, and its observed realisation $\bs{d}$, which corresponds to the measured data set used in the analysis.} Once the observed realisation $\bs{d}$ is fixed, the conditional probability density $p(\bs{D}=\bs{d}|\bs{\theta},M)$ is regarded as a function of the parameters, and is called the \textit{likelihood} function ${\cal L}$:
\begin{equation}
    {\cal L}(\bs{\theta}; \bs{d},M) \equiv p(\bs{d}|\bs{\theta},M).
\end{equation}In what follows, we will suppress the semicolon notation that emphasizes the dependence of the function ${\cal L}(\bs{\theta})$ on the fixed $\bs{d}$ and $M$.

\subsubsection{Likelihood}\label{sssec:likelihood}
We assume the likelihood of observing a data set $\bs{d}$ given our power spectrum and bispectrum models, $M(\bs{\theta})$, to be described by a multivariate Gaussian function
\begin{equation}\label{eq:likelihood}
    {\cal L}(\bs{\theta}) \propto {\rm exp}\left[-\frac{1}{2}[\bs{d} - M(\bs{\theta})]_i\,\hat{{\sf C}}^{-1}_{ij}[\bs{d} - M(\bs{\theta})]_j\right],
\end{equation}where the inverse covariance matrix $\hat{{\sf C}}^{-1}$ (see Section \ref{ssec:cov} for the construction of ${\sf C}$) is unbiased by using the correction derived from inverse-Wishart statistics \citep[see][and references therein]{hartlap2006}:
\begin{equation}
    \hat{{\sf C}}^{-1} = \frac{N_s - N_d - 2}{N_s-1}\, {\sf C}^{-1},
\end{equation}where $N_s = 1000$ is the total number of samples used to estimate ${\sf C}$, and $N_d = 72$ for the $P_0+P_2$, and $N_d = 108$ when including $B_{000}$ (each statistic has 36 $k$-bins over the scales we considered).

\subsubsection{Priors} \label{sssec:priors}
A key aspect of Bayesian analysis is the possibility of including \textit{prior} knowledge about $\bs{\theta}$ in the estimation process. For that, we must assume a prior probability distribution function $\pi(\bs{\theta})\equiv p(\bs{\theta},M)$. To obtain the final posterior for the parameters of interest $\bs{\theta}_*$, one needs to perform a marginalisation over the set of nuisance parameters $\bs{\psi}\in\mathbb{R}^n$:
\begin{equation}
    p(\bs{\theta}_*|\bs{d},M) = \int{\rm d}^n\psi\,\, {\cal L}(\bs{\theta})\,\pi(\bs{\theta}_*,\bs{\psi}).
\end{equation}

For illustrative purposes, let us consider the simplified case in which the priors are separated into contributions from the parameters of interest and the nuisance parameters, $\pi(\bs{\theta}_*,\bs{\psi})=\pi(\bs{\theta}_*)\pi(\psi)$, assuming a single $\psi$ and flat priors.

Suppose that the likelihood depends on $\psi$ only through its combination with a cosmology-dependent amplitude $A(\bs{\theta}_*)$ (which is the case of many EFT and bias parameters), that is, ${\cal L}(\bs{\theta}) \equiv {\cal L}(\psi\, A)$. Through the change of variables $x \equiv \psi \,A(\bs{\theta}_*)$, the marginalised posterior becomes
\begin{equation}
    p(\bs{\theta}_*|\bs{d},M) \propto \frac{\pi(\bs{\theta}_*)}{A(\bs{\theta}_*)} \int {\rm d}x\, {\cal L}(x).
\end{equation}

Although the data constrain only the combination $x = \psi\, A$, a flat prior on $\psi$ implies that the total prior space leading to the same preferred $x$ is larger for smaller $A$. In other words, this leads to a spurious preference in the marginalised posterior for small values of $A$ that is not driven by the data, but by the chosen parametrisation of the nuisance parameters.

To mitigate these effects, the authors of \cite{maus2025, tsedrik2026} proposed a reparametrisation of the nuisance parameters such that the priors are over $\tilde{\psi} = \psi\, A(\bs{\theta}_*)$, where the amplitudes are $\sigma_8$, and the Alcock–Paczy\'{n}ski factor $A_{\rm AP} \equiv (q_\parallel q_\perp^2)^{-1}$. With this new set of nuisance parameters, the volume of the $\tilde{\psi}$-space that fits the data does not change as we vary $\bs{\theta}_*$.

As an example, since $P^{(s)}_g(k,\mu)\propto A_{\rm AP} \,b_1^2\, P_{\rm lin}(k)$, and $P_{\rm lin}\sim\sigma_8^2$ (or equivalently $A_s$), instead of sampling directly over $b_1$, we consider the reparametrised linear bias $\tilde{b}_1 = b_1 \sigma_8 \sqrt{A_{\rm AP}}$. In what follows, we will adopt this reparametrisation for the biases and counterterms. In Appendix~\ref{app:validation}, we present the effect of this reparametrisation on the posterior distribution obtained from the mean of 25 LRG mocks, quantifying its impact on the parameter constraints. 

We quote the priors adopted in this work in Table~\ref{tab:priors}, with the explicit connection between the reparametrised nuisance parameters and the theoretical ones. For the tidal and third-order biases, we have chosen Gaussian priors centred at their respective co-evolution relations \cite{moradinezhad2021}. For that, we fix $b_1^{\rm ref}(z_{\rm eff})$ using the DESI DR1 parametrisation of the linear bias evolution, $b_1(z)=a(1+z)^2 + b$, where the coefficients $a$ and $b$ are calibrated from the unblinded DESI samples \cite{chaussidon2025}. This gives $b_1^{\rm ref}(z_{\rm eff}) = (1.891, 2.023, 2.184, 2.233)$ for the LRG1, LRG2, LRG3, and QSO samples, respectively. 

We adopt broad Gaussian priors for the counterterms, and slightly more modest Gaussian priors for the shot-noise amplitudes. As shall be discussed next, in Section~\ref{sssec:anamarg}, the choice of Gaussian priors allows for the analytical marginalisation of nuisance parameters, which reduces the dimensionality of the sampled parameter space and improves the efficiency and convergence of the sampler. This is particularly useful in the joint analyses where multiple tracers are combined and the number of sampled parameters increases.  

\begin{table}[htpb]
\centering
\caption{Priors adopted in this work. ``Reference'' priors are used for the results with DESI DR1 data, whereas the ``FolpsD-inspired'' priors are used only in mock tests to assess the robustness of the inferred constraints to the prior choices. We adopt two standard distributions for the priors: ${\cal U}_{[a,b]}\equiv{\cal U}(a,b)$ is the uniform distribution with bounds defined by the closed support $[a,b]$, and ${\cal N}(\mu,\sigma^2)$ is the Gaussian distribution centred at $\mu$ with variance $\sigma^2$. For the co-evolution relations of $b_s$ and $b_3$, we fixed $b_1^{\rm ref} = (1.891, 2.023, 2.184, 2.233)$ for the LRG1, LRG2, LRG3, and QSO samples, respectively. The power-spectrum counterterms ($c_0,\,c_2,\,c_4$) are related to the \texttt{folpsD} counterterms ($\alpha_0,\, \alpha_2,\,\alpha_4$) via the relations $\alpha_0=b_1^2\,c_0,\, \alpha_2=f\,b_1(c_0+c_2),\, \alpha_4=f^2\,c_2+f\,b_1\,c_4$. Shot-noise amplitudes are in units of $\bar{n}_g^{-1}$.}
\renewcommand{\arraystretch}{1.2}
\begin{tabular}{r@{$\,=\,$}lcc}
\multicolumn{4}{c}{${}$}\\
\toprule
\multicolumn{2}{c}{\textbf{Parameter}} & \multicolumn{2}{c}{\textbf{Priors}}\\
\midrule
\multicolumn{2}{c}{} & \textbf{Reference} & \textbf{FolpsD-inspired} \\
\midrule
\multicolumn{4}{c}{\textbf{Cosmological parameters}} \\
\midrule
\multicolumn{2}{c}{$\omega_{\rm cdm}$} & \multicolumn{2}{c}{${\cal U}(0.05,0.2)$}\\  
\multicolumn{2}{c}{$\omega_b$} & \multicolumn{2}{c}{${\cal N}(0.02218, 0.00055^2)$} \\
\multicolumn{2}{c}{$h$} & \multicolumn{2}{c}{${\cal U} (0.5,0.9)$} \\ 
\multicolumn{2}{c}{$n_s$} & \multicolumn{2}{c}{${\cal N}(0.9649, 0.042^2)$} \\
\multicolumn{2}{c}{$\ln(10^{10} A_s)$} & \multicolumn{2}{c}{${\cal U} (2,4)$} \\
\midrule
\multicolumn{4}{c}{\textbf{Galaxy bias parameters}} \\
\midrule
$\tilde{b}_1$ & $b_1\,\sigma_8\,\sqrt{A_{\rm AP}}$ & \multicolumn{2}{c}{${\cal U}(0.1,4)$} \\ 
$\tilde{b}_2$ & $b_2\,\sigma_8^2\,\sqrt{A_{\rm AP}}$ & \multicolumn{2}{c}{${\cal N}(0, 5^2)$} \\ 
$\tilde{b}_{K^2}$ & $b_{K^2}\,\sigma_8^2\,\sqrt{A_{\rm AP}}$ & \multicolumn{2}{c}{${\cal N}\left(-\frac{2}{7}(b_1^{\rm ref}-1), 20^2\right)$} \\
$\tilde{b}_{\rm td}$ & $b_{\rm td}\,\sigma_8^4\,A_{\rm AP}$ & \multicolumn{2}{c}{${\cal N}\left(\frac{23}{42}(b_1^{\rm ref} - 1), 1^2\right)$} \\
\midrule
\multicolumn{4}{c}{\textbf{Power spectrum counterterms}} \\
\midrule
$\tilde{c}_0$ & $c_0\,\sigma_8^2\,A_{\rm AP}$ & ${\cal N}(0, 50^2)$ & ${\cal N}(0, 12.5^2)$ \\
$\tilde{c}_2$ & $c_2\,\sigma_8^2\,A_{\rm AP}$ & ${\cal N}(0, 50^2)$ & ${\cal N}(0, 12.5^2)$ \\
$\tilde{c}_4$ & $c_4\,\sigma_8^2\,A_{\rm AP}$ & ${\cal N}(0, 50^2)$ & ${\cal N}(0, 12.5^2)$ \\
$\tilde{a}_0$ & $a_0\,A_{\rm AP}$ & \multicolumn{2}{c}{${\cal N}(0, 2^2)$} \\
$\tilde{a}_2$ & $a_2\,A_{\rm AP}/(f_{\rm sat}\sigma_v^2)$ & \multicolumn{2}{c}{${\cal N}(0, 5^2)$} \\
\midrule
\multicolumn{4}{c}{\textbf{Bispectrum counterterms}} \\
\midrule
$\tilde{c}_1$ & $c_1\,\sigma_8^2\,A_{\rm AP}$ & \multicolumn{2}{c}{${\cal N}(0, 20^2)$} \\
$\tilde{P}_{\rm shot}$ & $P_{\rm shot}\,A_{\rm AP}$ & \multicolumn{2}{c}{${\cal N}(0, 1^2)$} \\
$\tilde{B}_{\rm shot}$ & $B_{\rm shot}\,A_{\rm AP}$ & \multicolumn{2}{c}{${\cal N}(0, 1^2)$} \\
\bottomrule
\multicolumn{4}{c}{}
\label{tab:priors}
\end{tabular}
\end{table}

\subsubsection{Analytical marginalisation} \label{sssec:anamarg}

To improve the efficiency of posterior sampling, especially in the multiple-tracer analyses, we consider the approach proposed in \cite{damico2019} and perform analytical marginalisation of nuisance parameters $\bs{\psi}_{\rm lin}$ entering linearly in the EFT power-spectrum model. For that, we consider the theory as 
\begin{equation}\label{eq:anamarg}
    P(k; \bs{\theta}_*,\bs{\psi}) = P_0(k; \bs{\theta}_*,\bs{\psi}_{\rm nl})+ \sum_i \psi_{{\rm lin},i}\, P_i(k; \bs{\theta}_*,\bs{\psi}_{\rm nl}),
\end{equation}where $P_0(\bs{\theta}_*,\bs{\psi}_{\rm nl}) = P(\bs{\theta}_*,\bs{\psi}_{\rm nl}, \bs{\psi}_{\rm lin} = 0)$ is the constant part of the theory (i.e., independent of $\bs{\psi}_{\rm lin}$), and 
\begin{equation}
    P_i(\bs{\theta}_*,\bs{\psi}_{\rm nl}) \equiv \frac{\partial P(\bs{\theta}_*,\bs{\psi})}{\partial\psi_{{\rm lin},i}}
\end{equation}is the response of the theory to the $i$-th linear nuisance parameter. For simplicity, we suppressed the multipole index of the EFT power spectrum $\ell$, and in what follows we will also suppress the $k$-mode dependence. In Equation~\eqref{eq:anamarg}, we separate $\bs{\psi}=(\bs{\psi}_{\rm lin}, \bs{\psi}_{\rm nl})$ into the nuisance parameters that enter the model linearly and non-linearly.

If the priors on $\psi_{{\rm lin},i}$ are all Gaussians ${\cal N}(\mu_i,\sigma_i^2)$, then the likelihood in Equation~\eqref{eq:likelihood} can be analytically integrated over those priors:
\begin{equation}
    \ln{\cal L}_{\rm marg}(\bs{\theta}) = -\frac{1}{2} F^{(0)} + \frac{1}{2}F^{(1)}_i\left(F^{(2)}_{ij}\right)^{-1}F^{(1)}_j - \frac{1}{2}\ln\left({\rm det}\,F^{(2)}\right),
\end{equation}
where
\begin{equation*}
    F_0 = \Delta_a\,\hat{{\sf C}}^{-1}_{ab}\,\Delta_b + \sum_i \frac{\mu_i^2}{\sigma_i^2},\quad F_{1,i} = - P_{i,a}\,\hat{{\sf C}}^{-1}_{ab}\,\Delta_b + \frac{\mu_i}{\sigma_i^2},\quad F_{2,ij} = P_{i,a}\,\hat{\sf C}^{-1}_{ab}\,P_{j,b} + \frac{\delta_{ij}}{\sigma_i^2},
\end{equation*}and $\bs{\Delta}$ are the residuals between $P_0$ and the data.

We marginalise over the set of power spectrum counterterms $\lbrace \tilde{c}_0, \tilde{c}_2, \tilde{c}_4, \tilde{a}_0, \tilde{a}_2 \rbrace$ for the multi-tracer cases, where the number of sampled parameters grows from 14 and 17 for the $P_0+P_2$ and $P_0+P_2+B_{000}$, to 41 and 53, respectively. For the single-tracer analyses, however, we do not perform any analytical marginalisation and directly sample over the full set of parameters.

\subsubsection{Sampler} \label{sssec:sampler}

We sample our posterior distribution with the {\tt pocoMC} sampler \cite{pocomc}: it implements the Preconditioned Monte Carlo (PMC) algorithm that combines the Sequential Monte Carlo (SMC) method with Normalising Flows (NF). The latter is used to map the posterior distribution in the parameter space $\theta$ into a new coordinate system $u = f^{-1}(\theta)$ in which the posterior geometry is simplified (more Gaussian and less correlated), making the sampling more efficient \cite{karamanis2022}.

SMC uses a set of particles that evolve through a sequence of distributions, from the prior to the posterior, as 
\begin{equation}
    p_t(\bs{\theta}) \propto \pi(\bs{\theta}) {\cal L}(\bs{\theta})^{\beta_t},
\end{equation}where $\beta_t \in [0,1]$. When $\beta = 1$, the full posterior is sampled. The sampler stops when $\beta > 0.9999$ and the ``Effective Sample Size'' (ESS) is ${\rm ESS}\geq n_{\rm tot}$, where $n_{\rm tot}$ is the total number of particles chosen to sample the posterior. Unless otherwise stated, we set $n_{\rm tot}=2\times10^4$ in this work.

\subsection{Emulators} \label{ssec:emulator}

A major computational bottleneck in modern cosmological inference is the repeated evaluation of accurate but expensive forward models within likelihood and sampling pipelines. Emulators address this problem by learning a fast surrogate of the target observable from a training set of high-fidelity calculations, thereby enabling orders-of-magnitude speed-ups while retaining sufficient accuracy for parameter estimation. This approach is now widely used in cosmology, from early CMB and likelihood emulation to recent emulators for cosmological power spectra, galaxy clustering, and EFT-based large-scale-structure predictions \citep{Fendt:2006uh,SpurioMancini:2021ppk,Bonici:2023xjk,Arico:2021izc,Mootoovaloo:2021rot,Donald-McCann:2021nxc,Bonici:2025ltp}.

For all of our $\Lambda$CDM analyses, we compute the linear matter power spectrum using the \texttt{baccoemu} emulator \citep{Arico:2021izc}, which predicts the linear power spectrum with sub-percent accuracy over the scales relevant for LSS analyses.

To explore models beyond $\Lambda$CDM, in particular the $w_0w_a$ analysis presented in Section~\ref{ssec:w0wa}, we employ instead the \texttt{Mapse.jl} emulator, as it was trained on a larger parameter space than \texttt{baccoemu}. It belongs to the same family as \texttt{Capse.jl} and \texttt{Effort.jl}, sharing the neural-network infrastructure implemented in \texttt{Abstract\-Cosmological\-Emulators.jl} \citep{Bonici:2023xjk,Bonici:2025ltp,Bonici:inprepMapse}. In this work we use its JAX implementation, \texttt{jaxmapse},\footnote{The code is publicly available at \url{https://github.com/CosmologicalEmulators/jaxmapse}.} which provides the version adopted in our $w_0w_a$ analysis. To reach an accuracy of about $0.1\%$ over a wide $w_0w_aM_\nu$CDM parameter space, \texttt{Mapse.jl} emulates only the transfer function, treats $\ln(10^{10}A_s)$ and $n_s$ analytically, factors out the leading redshift dependence through division by $D^2(z)$, removes additional smooth structure through a symbolic approximation obtained by genetic programming, and finally compresses the target space via principal component analysis before the regression stage \citep{Bonici:inprepMapse}. By doing this, \texttt{jaxmapse} reaches submillisecond performance when computing the linear power spectrum for a single redshift.

As a consistency test, we repeated the same $\Lambda$CDM inference with \texttt{jaxmapse} and \texttt{baccoemu} emulators. The shifts in the posterior centres are much smaller than the corresponding 1-$\sigma$ uncertainties for all parameters. The posterior widths also agree at the few-percent level, with the largest difference being an approximately $6\%$ tighter constraint on $\omega_{\rm cdm}$, when using \texttt{baccoemu}. At the precision of the present analysis, the two implementations therefore lead to statistically consistent cosmological constraints.

\section{Results} \label{sec:results}

In this section, we present the cosmological constraints obtained from the DESI DR1 LRG and QSO clustering measurements. We keep the maximum wavenumber $k_{\rm max} = 0.2\,\,h/{\rm Mpc}$ fixed in the main $P_0+P_2+B_{000}$ analyses, for all tracers and redshift bins. This choice is motivated by the validation presented in \cite{bansal2026} and by the cut-sky tests discussed in Appendix~\ref{app:validation}. For LRGs, \cite{bansal2026} found that the joint one-loop EFT power spectrum and tree-level bispectrum analysis remains unbiased up to $k_\text{max}\simeq 0.20 \,h/ \text{Mpc}$ even in the absence of phenomenological damping. Interestingly, the extended reach at $0.20\,h/\text{Mpc}$ does not appear to be primarily associated with the damping. The power-spectrum and bispectrum monopoles become strongly correlated towards smaller scales, with the $P_0$-$B_{000}$ correlation reaching values $\gtrsim0.5$ around $k \sim 0.1\,h/\text{Mpc}$ \cite{bansal2026}. This suggests that the additional bispectrum bins at these scales do not provide fully independent information and can therefore have a comparatively limited leverage on the joint cosmological posterior.  For QSOs, the EFT tests of \cite{bansal2026} favoured a somewhat more conservative scale, $k_\text{max}^{000}\simeq0.16\,h/{\rm Mpc}$, making the cut-sky validation presented in Appendix~\ref{app:validation} particularly important for assessing the use of $k_\text{max}=0.2\,h/\text{Mpc}$ in the present analysis. 

The median of all of our marginalised posteriors with the 1-$\sigma$ confidence intervals can be found in Table~\ref{tab:main_results}. For completeness, in Appendix~\ref{app:extra} we present the posteriors on the cosmological parameters for each individual tracer.

\subsection{Power-spectrum-only analysis}\label{ssec:powerspec-only}

Before we discuss the role of the bispectrum monopole in the analysis, we start by validating the statistical consistency of our power-spectrum-only analyses ($P_0+P_2$) with the DESI DR1 ``full-modelling'' results for the same individual tracers \cite{DR1Pk}. Since both analyses use the same measurements and covariance matrices, the posterior estimates can be correlated, so we do not assign a formal significance to shifts in the posteriors. Instead, we compare the constraints directly (see Table~\ref{tab:main_results} for numerical results), finding good agreement for all four tracers. 

\begin{figure}[ht]
    \centering
    \includegraphics[width=\linewidth]{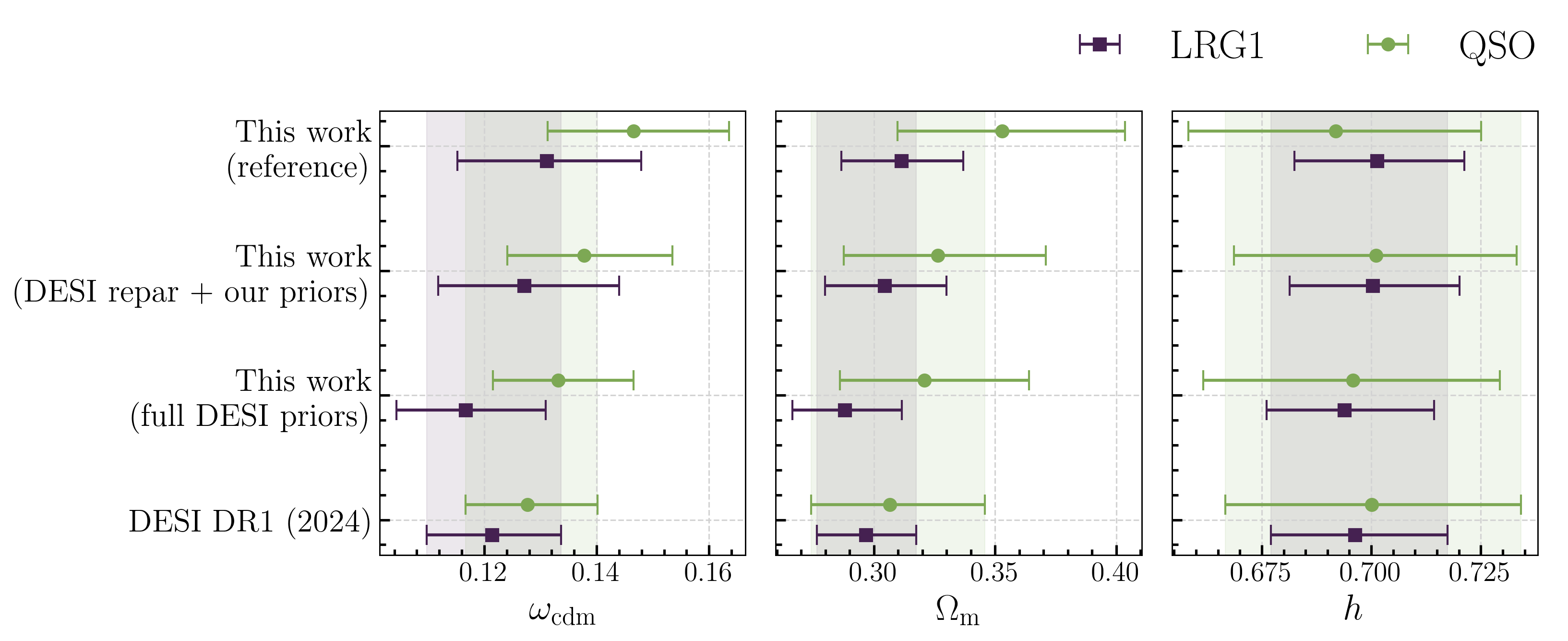}
    \vspace{-5mm}
    \caption{Comparison of $P(k)$-only constraints obtained in this work with different choices of nuisance-parameter priors and reparametrisation, and with the DESI DR1 results \cite{DR1Pk}. The points show the posterior medians for the LRG1 (squares) and QSO (circles) samples, with horizontal error bars indicating the marginalised 1-$\sigma$ errors. For guidance, the shaded bands show the corresponding DESI 2024 1-$\sigma$ intervals for both tracers.}
    \label{fig:DESI_comparison}
\end{figure}

The largest differences with respect to the DESI 2024 power-spectrum-only constraints are observed for the LRG1 and QSO samples, where the posterior centres of $\omega_{\rm cdm}$ are shifted to higher values. In Figure~\ref{fig:DESI_comparison}, we compare the DESI 2024 full-modelling results with our reference choice of reparametrisation and priors (see Table~\ref{tab:priors}), but we also show results obtained by removing the $A_{\rm AP}$ scaling and reparametrising only the biases with $\sigma_8$, as done in \cite{DR1Pk} (``DESI repar.'' in the figure), first retaining our broad power-spectrum counterterms priors and then imposing the tighter DESI priors (``full DESI priors'' in the figure). As we show in Appendix~\ref{app:validation}, under the $\Lambda$CDM model our reparametrisation affects $\omega_{\rm cdm}$ more strongly than the other cosmological parameters and can broaden its marginalised posterior. Tests on the mean mock measurements nevertheless recover the fiducial cosmology within the expected uncertainties. The displacement observed in the data should therefore be interpreted as the combined effect of the realised data vector and the nuisance-parameter prior geometry, rather than as evidence for a systematic bias of the pipeline.

For the DR1 LRG1 bin, we find the 1-$\sigma$ uncertainty on $\Omega_m$ to be $\sigma\simeq0.025$ for the reference analysis and $\sigma\simeq0.023$ when adopting the full DESI priors. These correspond to uncertainties that are 22\% and 10\% larger than those of \cite{DR1Pk}, respectively. For the QSOs, we obtained $\sigma\simeq0.047$ (reference), $\sigma\simeq0.039$ (full DESI priors), larger than the results in \cite{DR1Pk} by 31\% and 9\%, respectively. We find the width of the $\omega_{\rm cdm}$ posterior to be affected by the prior adopted for the power-spectrum monopole $c_0$ counterterm: broader priors lead to broader posteriors for $\omega_{\rm cdm}$, while the remaining cosmological parameters are only weakly affected. Broader priors on $c_0$ therefore allow $\omega_{\rm cdm}$ greater freedom to shift towards higher values.

Still, our results are statistically compatible with those presented in the official DESI power-spectrum-only analysis. We stress the fact that we do not expect our results to be numerically identical to those of \cite{DR1Pk}, since we use independent pipelines, with different modelling choices, nuisance parameters and reparametrisation. 

While the single tracer analyses probe different survey volumes, redshift ranges and galaxy populations, each characterised by different bias parameters and degeneracy directions, combining them increases the statistical power not only through a larger effective volume, but also by helping to break parameter degeneracies present in the individual analyses. The different degeneracy directions of the LRG and QSO samples allow the multi-tracer analysis to break parameter degeneracies that remain in the individual fits, leading to tighter cosmological constraints.

Combining the three LRG redshift bins strengthens the power-spectrum-only constraints relative to those obtained from each bin separately: the uncertainty on $h$ decreases from $0.017$-$0.022$ in the individual bins to $\pm 0.012$ in the combined fit, corresponding to an improvement of about $30$-$50\%$. Similarly, the uncertainties on $\omega_{\rm cdm}$ are reduced by $30$-$45\%$, while the constraint on $\ln(10^{10}A_s)$ improves by $\sim40\%$. On the other hand, we only obtain a $7$-$10\%$ gain in $n_s$, with its uncertainty decreasing from 0.038-0.039 to 0.035. 

\begin{figure}[htpb]
    \centering
    \includegraphics[width=\linewidth]{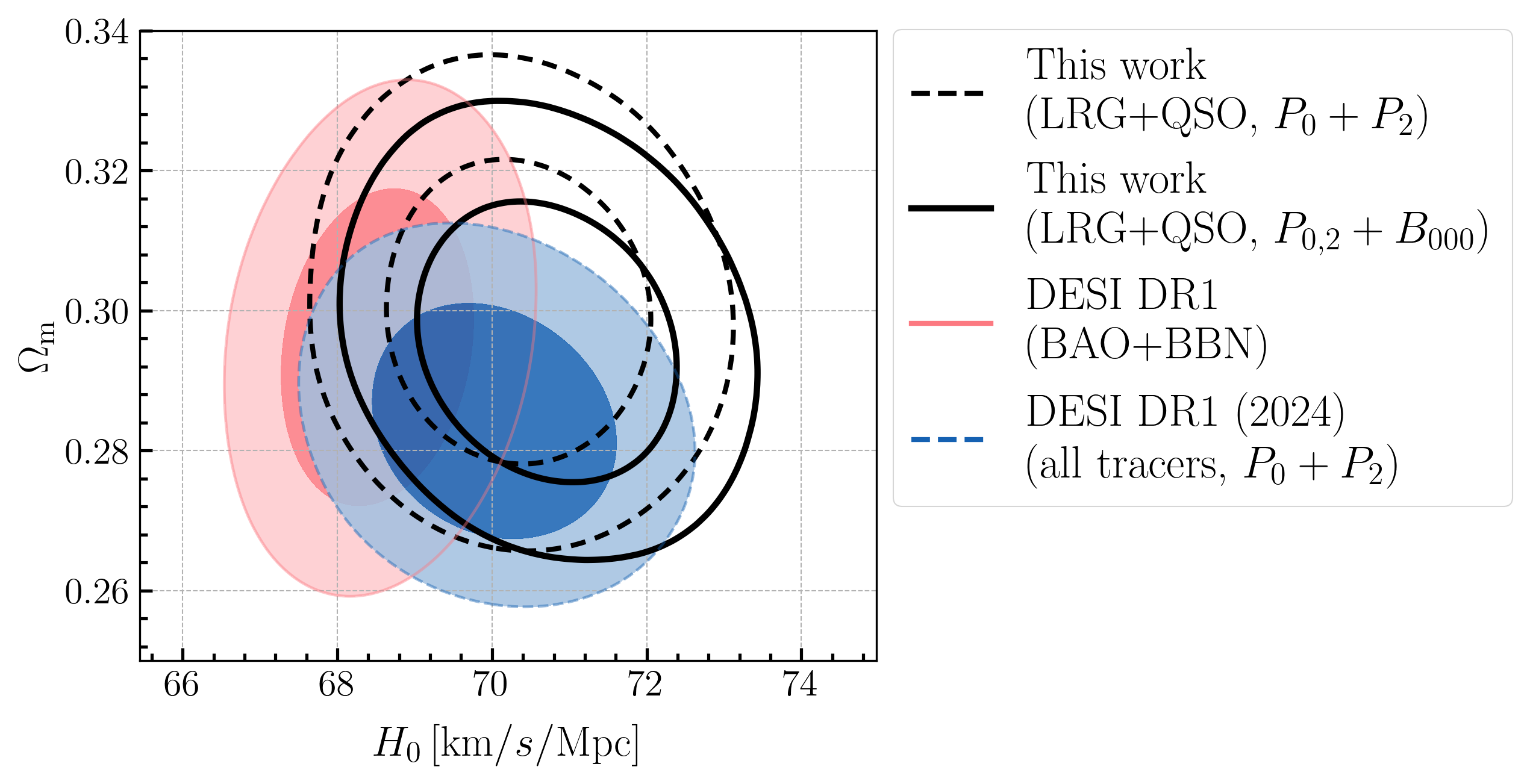}
    \caption{Comparison of the joint constraints on $\Omega_{\rm m}$ and $H_0$ obtained in this work with previous DESI DR1 analyses. Dashed and solid black contours show the constraints obtained from our joint LRG+QSO analysis using the power-spectrum monopole and quadrupole, $P_0+P_2$, and after additionally including the bispectrum monopole, $B_{000}$, respectively. We compare these to the DESI DR1 BAO+BBN constraints (solid contour) and the DESI ``full-modelling'' using all tracers available in that analysis (Bright Galaxy Sample and Emission Line Galaxies in addition to the three LRG bins and QSOs we considered).}
    \label{fig:Omega_mH0_comp}
\end{figure}

In Figure~\ref{fig:Omega_mH0_comp}, we further compared the joint LRG+QSO constraints on $\Omega_{\rm m}$ and $H_0$ with (\textit{i}) the DESI DR1 Baryon Acoustic Oscillations (BAO) + Big Bang Nucleosynthesis (BBN) priors \cite{desiBAO2025}, which are the same ones assumed in our analysis, and (\textit{ii}) the DESI full-shape results. Even though the individual LRG1 and QSO samples favour slightly higher $\Omega_{\rm m}$ values, as previously discussed, the combined constraints are mostly driven by the LRG3 sample, which provides the tightest cosmological constraints. 

Finally, LRGs and QSOs are complementary not only in terms of different parameter degeneracies: although QSOs have a larger covariance matrix than the individual LRG bins owing to their lower number density and correspondingly larger shot noise, they probe higher redshifts $(0.8<z<2.1)$, providing complementary information on the growth of structure. While the QSO sample yields weaker constraints than the LRGs for most parameters, it provides competitive constraints on $\ln(10^{10} A_s)$, with $\sigma\simeq0.19$ compared to $\sigma\simeq0.25$ for the individual LRG bins. This can be understood from the $f\,\sigma_8$ combination probed by the redshift-space power-spectrum multipoles: since $f(z)$ approaches its matter-dominated limit, $f(z)\rightarrow 1$, over the QSO redshift range, the degeneracy between the growth rate and $\sigma_8$ is reduced, allowing the clustering amplitude, and consequently $A_s$, to be more directly constrained by the power spectrum alone. 

\begin{figure}[htpb]
    \centering
    \includegraphics[width=0.75\linewidth]{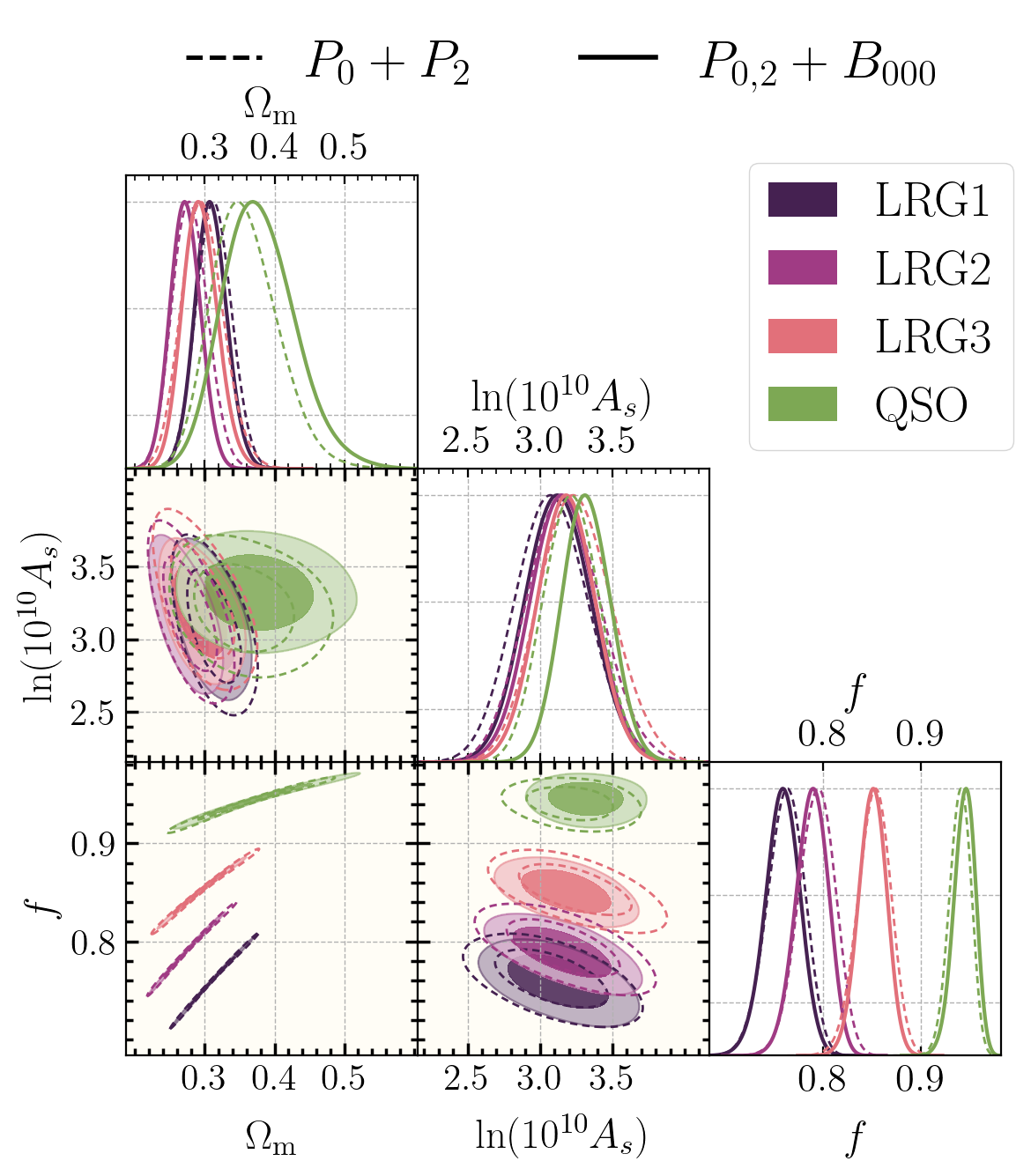}
    \caption{Marginalised posterior distributions for $\Omega_{\rm m}$, $\ln(10^{10}A_s)$, and the growth rate $f$, obtained from the individual LRG1, LRG2, LRG3, and QSO samples. Dashed contours show the constraints from the power-spectrum monopole and quadrupole, $P_0+P_2$, while solid contours include the bispectrum monopole, $P_0+P_2+B_{000}$. The filled regions indicate the corresponding 1- and 2-$\sigma$ intervals. The comparison highlights the complementary information provided by $B_{000}$, as well as the different parameter degeneracies probed by the low-redshift LRG samples and the higher-redshift QSO sample.}
    \label{fig:fAs}
\end{figure}

\subsection{Impact of the bispectrum}\label{ssec:bispectrum-analysis}

We now compare the impact of adding the bispectrum to the analyses with the power-spectrum-only cases. The role of the bispectrum \textit{quadrupole} was extensively discussed in \cite{bansal2026} (see their Appendix A for the case of the LRG2 mocks). However, for completeness, we provide a similar discussion with the DR1 data, focusing on the LRG1 bin as a proof of concept, and fixing $k_{\rm max} = 0.08\,\,h/{\rm Mpc}$ for the quadrupole $B_{202}$. As one can see in the direct comparison between the $P_0+P_2+B_{000}$ and $P_0+P_2+B_{000}+B_{202}$ cases (see Table~\ref{tab:main_results}), $B_{202}$ provides only marginal gains, changing the marginalised uncertainties by only a few percent, while leaving the posterior medians essentially unchanged. Therefore, we conclude that for the parameters, configurations, and scales considered here, $B_{202}$  provides little information beyond $P_0+P_2+B_{000}$, and we omit it from the remaining analyses.

\begin{figure}[htpb]
    \centering
    \includegraphics[width=\linewidth]{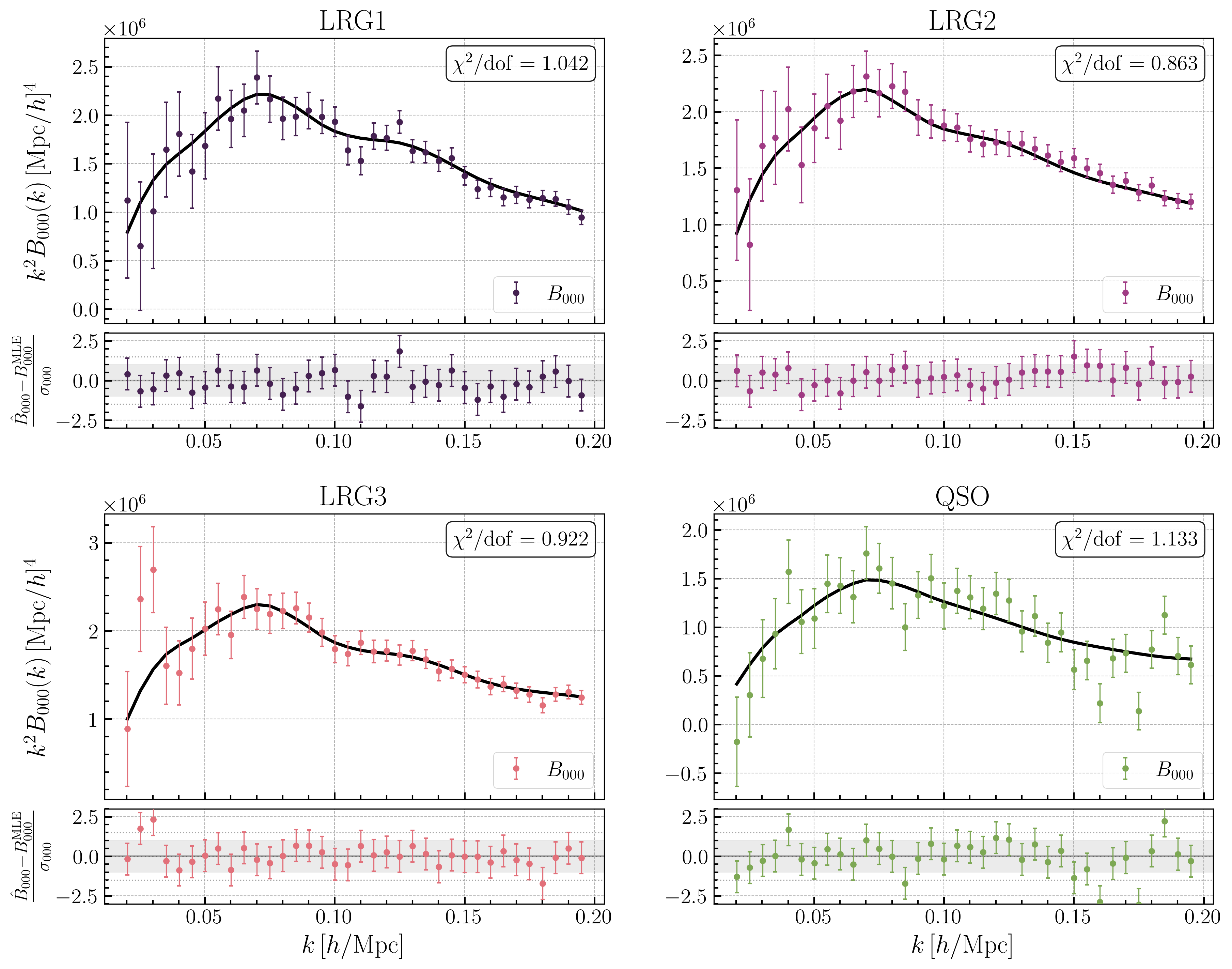}
    \caption{$B_{000}$ measurements and best-fitting predictions for the LRG1, LRG2, LRG3, and QSO samples. The upper panels show $k^2B_{000}(k)$, with coloured points representing the DESI DR1 measurements and black curves showing the window-convolved maximum-likelihood predictions $B_{000}^{\rm MLE} \equiv B_{000}(\hat{\bs{\theta}}_{\rm MLE})$ obtained separately from the corresponding single-tracer $P_0+P_2+B_{000}$ analyses. Error bars are derived from the diagonal of the bispectrum covariance. The lower panels show the residuals between the measurements and predictions, normalised by the corresponding standard deviation $\sigma_{000}$; dotted horizontal lines indicate the $\pm$1-$\sigma$ interval. The reduced $\chi^2$ for each bispectrum fit is reported in the upper-right corner.}
    \label{fig:bispectrum_fits}
\end{figure}

In Figure~\ref{fig:bispectrum_fits}, we show the bispectrum measurements with the maximum-likelihood model predictions. These were obtained separately for each tracer, from the parameter vector $\hat{\bs{\theta}}_{\rm MLE} \equiv \arg\max_{\bs{\theta}}{\cal L}(\bs{d}|\bs{\theta})$ that maximises the corresponding single-tracer $P_0+P_2+B_{000}$ likelihood (see the individual tracer constraints in Table~\ref{tab:main_results}). The model provided a good description of the four samples considered in this work, as can be seen by the reduced $\chi^2$ values.

Table~\ref{tab:bispectrum-gains} shows the change in the 1-$\sigma$ errors between the $P_0 + P_2$ ($\sigma_P$) and the $P_0 + P_2 + B_{000}$ ($\sigma_B$) fits, relative to $\sigma_P$. Adding $B_{000}$ improves the constraints on $\omega_{\rm cdm}$ and $\ln(10^{10}A_s)$ by 9-18\% and 8-20\%, respectively, in the single-tracer analyses. The corresponding reductions are 15\% and 10\% for the combined LRG sample. 

For comparison, \cite{forerosánchez2026} reports an approximately $26\%$ reduction in the uncertainty on $\sigma_8$ when including the LRG bispectrum in their joint DR1 clustering and DR2 BAO analysis, accompanied by an upward 1.1$\sigma$ shift in the posterior mean. Our combined LRG analysis yields a more modest, 5\% reduction in the $\sigma_8$ uncertainty. However, these gains are not directly comparable: their results include DR2 BAO information and employ the ShapeFit compression, a different bispectrum model, and deterministic coevolution relations for the tidal and third-order non-local biases. They also use a larger set of triangle shapes, whereas we only use the diagonal TripoSH configurations $B_{000}(k,k)$. Furthermore, their P+B analysis includes the power-spectrum hexadecapole $P_4$ and changes the scale cuts and binning relative to their $P$-only baseline analysis. Their quoted improvement therefore does not isolate the contribution of the bispectrum with the $P$-only data vector held fixed.

\begin{table}[ht]
\centering
\setlength{\tabcolsep}{5pt}
\caption{Percentage change in the 1-$\sigma$ uncertainties when adding the bispectrum monopole $B_{000}$ to the analysis: negative values indicate a reduction in the uncertainty, while positive values indicate an increase.}
\renewcommand{\arraystretch}{1.2}
\begin{tabular}{lcccc|cc}
{}\\
\toprule
\textbf{Parameter} & \textbf{LRG1} & \textbf{LRG2} & \textbf{LRG3} & \textbf{QSO} & \textbf{LRG (combined)} & \textbf{All tracers} \\
\midrule
$h$
& $-8\%$ & $-14\%$ & $-9\%$ & $+10\%$ & $+4\%$ & $0\%$ \\
$\omega_{\rm cdm}$
& $-18\%$ & $-14\%$ & $-17\%$ & $-9\%$ & $-15\%$ & $-6\%$ \\
$\ln(10^{10} A_s)$
& $-8\%$ & $-14\%$ & $-20\%$ & $-13\%$ & $-10\%$ & $-4\%$ \\
$n_s$
& $0\%$ & $-4\%$ & $-3\%$ & $0\%$ & $+4\%$ & $+8\%$ \\
\midrule
$\Omega_{\rm m}$
& $-12\%$ & $-9\%$ & $-17\%$ & $+9\%$ & $-3\%$ & $-4\%$ \\
$\sigma_8$
& $-2\%$ & $-12\%$ & $-14\%$ & $+1\%$ & $-5\%$ & $-1\%$ \\
\bottomrule
\end{tabular}
\label{tab:bispectrum-gains}
\end{table}

The different degeneracy directions probed by the individual tracers, and the corresponding tracer-dependent impact of $B_{000}$, are illustrated in Figure~\ref{fig:fAs} for $\Omega_{\rm m}$, $\ln(10^{10}A_s)$, and $f$. The impact of adding $B_{000}$ differs not only among the individual tracers, but also between the individual and combined analyses. This is not unexpected: as shown in \cite{bansal2026}, an important part of the additional information provided by the bispectrum comes from its ability to better constrain the second-order bias parameters $b_2$ and $b_s$, thereby modifying the degeneracies between galaxy bias and cosmology. For an individual tracer, these degeneracies can be relatively strong, so adding the bispectrum can lead to a sizeable improvement in several cosmological parameters. When different redshift bins and tracers are combined, however, the power spectra themselves already provide complementary information because the samples have different biases, redshifts, number densities and cosmological degeneracy directions. Part of the information provided by the bispectrum, for a single tracer, can therefore become redundant in the combined analysis, and the relative gain from adding $B_{000}$ need not follow the pattern seen for the individual samples.

\begin{figure}[H]
    \centering
    \includegraphics[width=0.85\linewidth]{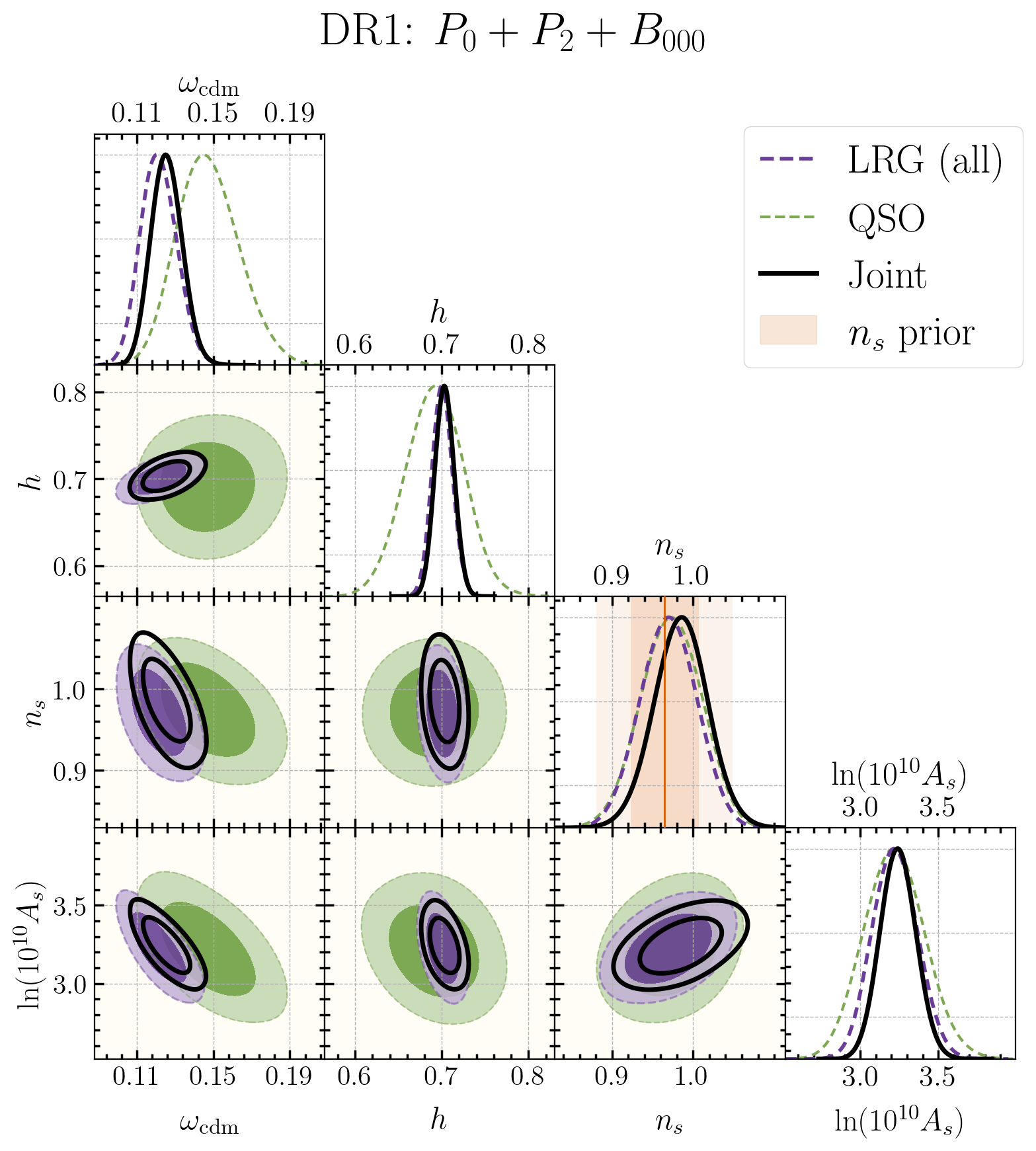}
    \caption{Posterior constraints on the cosmological parameters obtained from the DESI DR1 data, using $P_0$, $P_2$ and $B_{000}$. The joint constraint from all three LRG redshift bins and the QSO sample is shown in black. The 1-$\sigma$ values are quoted in Table~\ref{tab:main_results}.}
    \label{fig:finalresults_PkBk}
\end{figure}

Figure~\ref{fig:finalresults_PkBk} summarises the resulting $P_0+P_2+B_{000}$ constraints from the combined LRG sample, the QSO sample, and their joint analysis. Although the joint $n_s$ posterior is shifted towards slightly higher values than the individual-tracer posteriors, it remains within the 1-$\sigma$ region of the adopted Gaussian prior (see Table~\ref{tab:priors}), which is shown in the shaded beige region of Figure~\ref{fig:finalresults_PkBk}. The joint posterior, which gives $n_s=0.988\pm0.035$, is displaced from the prior mean by $\lesssim 0.7\sigma$. By adding $B_{000}$, the posterior median is shifted by less than $0.1\sigma$. We therefore find no statistically significant preference for a different value of $n_s$, and do not assign a physical interpretation to this shift. 

{
\captionsetup{font=normalsize}
\footnotesize
\setlength{\LTcapwidth}{\linewidth}
\setlength{\tabcolsep}{3.5pt}
\renewcommand{\arraystretch}{1.2}
\begin{longtable}{lcccc|cc}
\caption{Median of the marginalised posteriors, with 1-$\sigma$ constraints. We do not quote $\omega_{\rm b}$ values, as they are prior dominated. DESI 2024 constraints were obtained directly from the chains of \cite{DR1Pk}. For the LRG1 analysis, we also show the results obtained by including $B_{202}$ (see discussion in Section~\ref{ssec:bispectrum-analysis}).}
\label{tab:main_results}
{}\\
\toprule
\multicolumn{7}{c}{\textbf{LRG1}}\\
\midrule
& $h$ & $\omega_{\rm cdm}$ & $\ln(10^{10}A_s)$ & $n_s$ & $\Omega_{\rm m}$ & $\sigma_8$\\
\midrule
$P_0 + P_2$ (DESI 2024)
& $0.696^{+0.021}_{-0.019}$
& $0.121^{+0.012}_{-0.012}$
& $3.08^{+0.23}_{-0.23}$
& $0.968^{+0.040}_{-0.040}$
& $0.297^{+0.021}_{-0.020}$
& $0.842^{+0.089}_{-0.087}$\\
$P_0 + P_2$
& $0.701^{+0.020}_{-0.019}$
& $0.131^{+0.017}_{-0.016}$
& $3.08^{+0.25}_{-0.24}$
& $0.958^{+0.037}_{-0.038}$
& $0.313^{+0.026}_{-0.025}$
& $0.885^{+0.084}_{-0.080}$\\
$P_0 + P_2 + B_{000}$
& $0.700^{+0.019}_{-0.017}$
& $0.129^{+0.014}_{-0.013}$
& $3.13^{+0.23}_{-0.22}$
& $0.966^{+0.037}_{-0.038}$
& $0.308^{+0.023}_{-0.022}$
& $0.898^{+0.082}_{-0.079}$\\
$P_0 + P_2 + B_{000} + B_{202}$
& $0.698^{+0.019}_{-0.017}$
& $0.127^{+0.015}_{-0.014}$
& $3.11^{+0.24}_{-0.23}$
& $0.961^{+0.039}_{-0.037}$
& $0.307^{+0.024}_{-0.023}$
& $0.881^{+0.082}_{-0.075}$\\
\toprule
\multicolumn{7}{c}{\textbf{LRG2}}\\
\midrule
& $h$ & $\omega_{\rm cdm}$ & $\ln(10^{10}A_s)$ & $n_s$ & $\Omega_{\rm m}$ & $\sigma_8$\\
\midrule
$P_0 + P_2$ (DESI 2024)
& $0.731^{+0.022}_{-0.021}$
& $0.127^{+0.013}_{-0.013}$
& $3.09^{+0.23}_{-0.21}$
& $0.979^{+0.040}_{-0.040}$
& $0.279^{+0.021}_{-0.020}$
& $0.884^{+0.080}_{-0.073}$\\
$P_0 + P_2$
& $0.738^{+0.022}_{-0.022}$
& $0.129^{+0.019}_{-0.017}$
& $3.16^{+0.26}_{-0.25}$
& $0.970^{+0.039}_{-0.040}$
& $0.279^{+0.023}_{-0.024}$
& $0.926^{+0.082}_{-0.076}$\\
$P_0 + P_2 + B_{000}$
& $0.747^{+0.020}_{-0.018}$
& $0.130^{+0.016}_{-0.015}$
& $3.16^{+0.22}_{-0.22}$
& $0.964^{+0.039}_{-0.037}$
& $0.272^{+0.022}_{-0.021}$ 
& $0.925^{+0.072}_{-0.067}$ \\
\toprule
\multicolumn{7}{c}{\textbf{LRG3}}\\
\midrule
& $h$ & $\omega_{\rm cdm}$ & $\ln(10^{10}A_s)$ & $n_s$ & $\Omega_{\rm m}$ & $\sigma_8$\\
\midrule
$P_0 + P_2$ (DESI 2024)
& $0.679^{+0.018}_{-0.017}$
& $0.113^{+0.012}_{-0.012}$
& $3.10^{+0.24}_{-0.22}$
& $0.975^{+0.038}_{-0.038}$
& $0.293^{+0.025}_{-0.024}$
& $0.809^{+0.077}_{-0.072}$\\
$P_0 + P_2$
& $0.681^{+0.017}_{-0.017}$
& $0.114^{+0.015}_{-0.014}$
& $3.24^{+0.26}_{-0.24}$
& $0.964^{+0.039}_{-0.039}$
& $0.296^{+0.031}_{-0.029}$
& $0.877^{+0.084}_{-0.078}$\\
$P_0 + P_2 + B_{000}$
& $0.686^{+0.016}_{-0.015}$
& $0.115^{+0.013}_{-0.011}$
& $3.18^{+0.20}_{-0.20}$
& $0.965^{+0.038}_{-0.038}$
& $0.292^{+0.026}_{-0.024}$
& $0.856^{+0.071}_{-0.068}$\\
\toprule
\multicolumn{7}{c}{\textbf{QSO}}\\
\midrule
& $h$ & $\omega_{\rm cdm}$ & $\ln(10^{10}A_s)$ & $n_s$ & $\Omega_{\rm m}$ & $\sigma_8$\\
\midrule
$P_0 + P_2$ (DESI 2024)
& $0.700^{+0.034}_{-0.034}$
& $0.127^{+0.013}_{-0.011}$
& $3.24^{+0.19}_{-0.19}$
& $0.976^{+0.037}_{-0.037}$
& $0.307^{+0.039}_{-0.032}$
& $0.946^{+0.078}_{-0.073}$ \\ 
$P_0 + P_2$
& $0.692^{+0.033}_{-0.034}$
& $0.147^{+0.017}_{-0.015}$
& $3.22^{+0.19}_{-0.19}$
& $0.973^{+0.038}_{-0.037}$
& $0.354^{+0.051}_{-0.043}$
& $1.010^{+0.085}_{-0.072}$ \\
$P_0 + P_2 + B_{000}$
& $0.670^{+0.039}_{-0.035}$
& $0.146^{+0.015}_{-0.014}$
& $3.32^{+0.17}_{-0.16}$
& $0.973^{+0.038}_{-0.037}$
& $0.375^{+0.054}_{-0.048}$
& $1.053^{+0.084}_{-0.075}$ \\
\toprule
\multicolumn{7}{c}{\textbf{LRG (combined)}}\\
\midrule
& $h$ & $\omega_{\rm cdm}$ & $\ln(10^{10}A_s)$ & $n_s$ & $\Omega_{\rm m}$ & $\sigma_8$\\
\midrule
$P_0 + P_2$
& $0.700^{+0.012}_{-0.011}$
& $0.121^{+0.010}_{-0.010}$
& $3.22^{+0.15}_{-0.14}$
& $0.970^{+0.035}_{-0.035}$ 
& $0.294^{+0.016}_{-0.015}$ 
& $0.908^{+0.049}_{-0.045}$ \\
$P_0 + P_2 + B_{000}$
& $0.709^{+0.012}_{-0.012}$
& $0.124^{+0.009}_{-0.008}$
& $3.11^{+0.13}_{-0.13}$
& $0.964^{+0.036}_{-0.037}$ 
& $0.293^{+0.015}_{-0.015}$
& $0.874^{+0.045}_{-0.044}$ \\
\toprule
\multicolumn{7}{c}{\textbf{All tracers}}\\
\midrule
& $h$ & $\omega_{\rm cdm}$ & $\ln(10^{10}A_s)$ & $n_s$ & $\Omega_{\rm m}$ & $\sigma_8$\\
\midrule
$P_0 + P_2$
& $0.703^{+0.011}_{-0.011}$
& $0.125^{+0.008}_{-0.008}$
& $3.24^{+0.12}_{-0.11}$
& $0.985^{+0.032}_{-0.033}$
& $0.300^{+0.014}_{-0.014}$
& $0.943^{+0.040}_{-0.038}$\\
$P_0 + P_2 + B_{000}$
& $0.707^{+0.011}_{-0.011}$
& $0.125^{+0.008}_{-0.007}$
& $3.24^{+0.11}_{-0.11}$
& $0.988^{+0.035}_{-0.035}$
& $0.297^{+0.013}_{-0.014}$
& $0.944^{+0.039}_{-0.038}$\\
\bottomrule
\end{longtable}
}

\subsection[
  \texorpdfstring{$w_0w_a$CDM constraints}{w0waCDM constraints}
]{
  \texorpdfstring{$\bs{w_0w_a}$CDM constraints}{w0waCDM constraints}
}
\label{ssec:w0wa}

We extended the $\Lambda$CDM analysis by allowing a time-evolving dark-energy equation of state $w(a) = w_0+w_a(1-a)$ \cite{chevallier2001,linder2002}. As mentioned in Section~\ref{ssec:emulator}, we employed \texttt{jaxmapse} for the $w_0w_a$CDM analysis to allow for the prior $w_0\sim{\cal U}(-3,1)$, $w_a \sim {\cal U}(-3,2)$, with the constraint $w_0 + w_a <0$, which is broader than the one available with \texttt{baccoemu}.

\begin{figure}[htpb]
    \centering
    \includegraphics[width=0.75\linewidth]{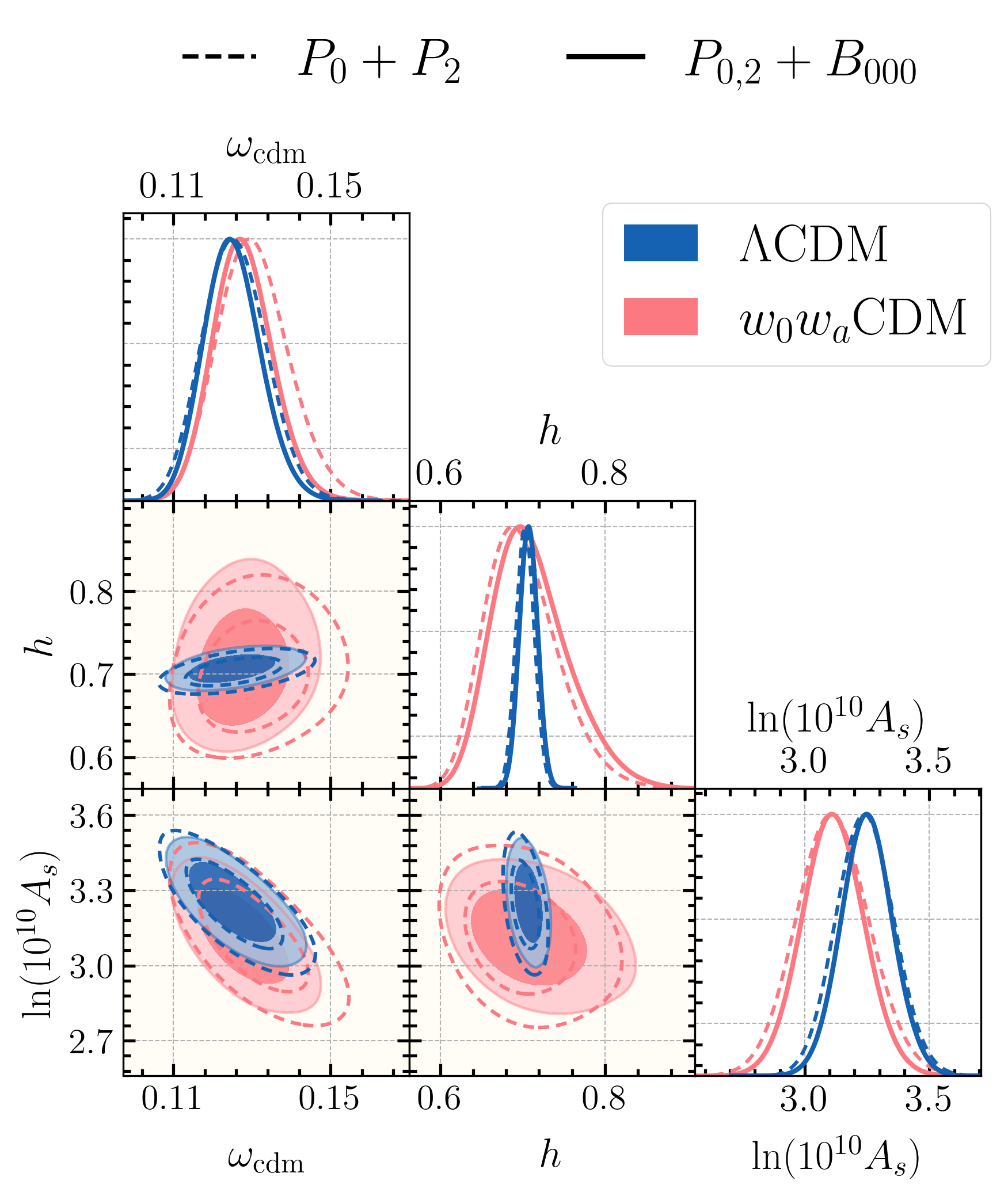}
    \caption{Posterior constraints obtained from a joint analysis of all DESI DR1 LRG bins and QSOs, assuming $\Lambda$CDM (blue) and $w_0w_a$CDM (coral) models. Constraints were obtained using the power-spectrum monopole and quadrupole, $P_0+P_2$ (dashed), and additionally including the bispectrum monopole $B_{000}$ (solid). Within the $w_0w_a$CDM model, the posterior distributions are broader than in the $\Lambda$CDM case due to the additional freedom introduced by $w_0$ and $w_a$.}
    \label{fig:w0wa_cosmopars}
\end{figure}

Figure~\ref{fig:w0wa_cosmopars} shows the effect of extending the dark-energy sector beyond a cosmological constant. Allowing the dark-energy equation of state to vary broadens the cosmological posteriors relative to the $\Lambda$CDM case, as expected from the additional freedom and the strong $w_0$-$w_a$ degeneracy. From the joint power-spectrum-only analysis we obtain
\begin{equation}
w_0=-0.73^{+0.36}_{-0.44}, \qquad
w_a=-1.25^{+1.37}_{-1.11},
\end{equation}
while the addition of $B_{000}$ gives
\begin{equation}
w_0=-0.74^{+0.34}_{-0.43}, \qquad
w_a=-1.40^{+1.30}_{-1.00}.
\end{equation}

The shifts are small compared with the marginalised uncertainties, and the bispectrum monopole does not significantly reduce the allowed $w_0$-$w_a$ region. The constraints remain compatible with a cosmological constant. For the data vector and scale cuts considered here, the main contribution of $B_{000}$ is therefore to the $\Lambda$CDM matter-density and amplitude parameters rather than to the time evolution of dark energy.

\section{Conclusions} \label{sec:conclusions}

In this work, we have presented a joint full-shape analysis of the DESI DR1 power spectrum and bispectrum for three LRG redshift bins and the QSO sample. We combined a one-loop EFT description of the power spectrum with the tree-level bispectrum, decomposed in the TripoSH basis, and consistently propagated the DESI survey geometry through the window convolution of both statistics. Building on the TripoSH window formalism of \cite{wang2025}, the bispectrum convolution can be implemented as a linear transformation of the theoretical multipoles, which makes its repeated evaluation in the likelihood inexpensive. We have tested this procedure on cut-sky mocks and find that it reproduces the measured bispectrum multipoles over the range of scales used in the analysis (Figure~\ref{fig:convolution_test}). Our power-spectrum-only constraints are in good agreement with both the DESI full-shape analysis and the independent BAO+BBN constraints (Figure~\ref{fig:Omega_mH0_comp}). The largest differences are found for $\omega_{\rm cdm}$ in the individual LRG1 and QSO fits, where the posterior is particularly sensitive to the counterterm priors and to the nuisance-parameter reparametrisation. Tests on the mean cut-sky mocks nevertheless recover the fiducial cosmology within the expected uncertainties.

Adding the bispectrum monopole changes the constraints in a way that depends strongly on the tracer and on the cosmological parameter. The largest and most persistent gains are found for $\omega_{\rm cdm}$ and $\ln(10^{10}A_s)$, with reductions of up to 18\% and 20\% in the marginalised constraints, respectively, in the single-tracer analysis. The improvement varies across the individual redshift bins and becomes different once the samples are combined (see Table \ref{tab:bispectrum-gains}). This behaviour is consistent with the picture found in \cite{bansal2026}: the bispectrum, decomposed into TripoSH multipoles, constrains the second-order bias parameters $b_2$ and $b_s$, which are only weakly constrained by the power spectrum, and therefore breaks part of the degeneracy between galaxy bias and cosmology. For an individual tracer, this can translate into a sizeable improvement in several cosmological parameters. When different redshift bins and tracers are combined, however, their power spectra already carry complementary information through their different biases, redshifts, number densities and parameter degeneracies, so the relative gain from adding $B_{000}$ need not follow the pattern found for the individual samples. We have also tested the bispectrum quadrupole $B_{202}$ for the LRG1 sample and $k_{\rm max} = 0.08\,\,h/{\rm Mpc}$, finding that it only produces marginal changes in the cosmological constraints, and we therefore find that most of the additional bispectrum information in the present analysis is already contained in $B_{000}$. However, this does not exclude possible gains from adding the $B_{202}$ quadrupole, or other possible multipoles in this basis \cite{byun2022}.

We have further explored whether the additional information from the bispectrum translates into stronger constraints when the dark-energy equation of state is allowed to evolve. In the $w_0w_a$CDM model, the extra freedom substantially broadens the cosmological posteriors, and the addition of $B_{000}$ produces only a mild change in the allowed $w_0$-$w_a$ region. The resulting posteriors remain compatible with a cosmological constant, although the posterior reaches the prior boundaries in the dark-energy parameter space. Thus, with the present data vector and scale cuts, the main benefit of the bispectrum is in the determination of the matter-sector and clustering-amplitude parameters rather than in constraining the time evolution of dark energy.

An important aspect of the analysis is the use of the bispectrum monopole up to $k_\text{max}=0.2\,h/\text{Mpc}$ despite employing a tree-level model. For LRGs, this choice is supported by the periodic-box tests of \cite{bansal2026}, where the joint one-loop EFT power spectrum and tree-level bispectrum analysis remained unbiased up to approximately this scale even without phenomenological damping. That work also found that the cross-correlation between $P_0$ and $B_{000}$ becomes large towards smaller scales, suggesting that the additional bispectrum bins are not providing completely independent information to the joint likelihood. The cut-sky tests presented here provide the corresponding validation in the presence of the DESI survey geometry and are particularly relevant for the QSO sample. This empirical scale reach should not be interpreted as a statement that the perturbative accuracy of the tree-level bispectrum itself extends to arbitrarily small scales. The present analysis is restricted to diagonal bispectrum configurations and to the LRG and QSO samples, with the bispectrum model kept at tree level. Extending the analysis to a larger set of triangle configurations, additional tracers, in particular ELGs, and higher-order bispectrum modelling will be important for determining how much more information can be extracted from the higher-order clustering of future DESI data.

\dataavailability
The data used in this work are available from the corresponding author upon reasonable request.

\acknowledgements

CG would like to thank Guilherme Brando and Henrique Rubira for useful discussions on all sorts of topics and, for numerous helpful conversations during the initial stages of this work, Thiago Mergulhão and Richard Neveux. Many thanks to Minas Karamanis for all the helpful comments and guidance with \texttt{pocoMC}, Eduardo Vitral for plotting inspiration and feedback and, finally, \textit{cheers} to Danny Horta for the encouragement during the final stages of this work. AA acknowledges financial support from SECIHTI grants CBF2023-2024-162 and  CBF-2025-I-2795, and grant DGAPA-PAPIIT IA101825. CG and FB acknowledge the support of the Royal Society through the University Research Fellowship.\\

\noindent We made extensive use of the \href{https://github.com/lesgourg/class_public}{\texttt{CLASS}} code \citep{blas2011}, and the following Python packages and libraries: \href{https://numpy.org/}{\texttt{numpy}} \citep{harris2020}, \href{https://scipy.org/}{\texttt{scipy}} \citep{2020SciPy}, \href{https://matplotlib.org/}{\texttt{matplotlib}} \citep{hunter2007}, and \href{https://getdist.readthedocs.io/en/latest/}{\texttt{GetDist}} \cite{getdist}. The AI technologies \href{https://chat.openai.com/}{\textsc{ChatGPT}} and \href{https://chat.deepseek.com/}{\textsc{DSeek}} were used to improve readability of the manuscript and the main pipeline used for analysis (including debugging and enhancing its efficiency). The authors take full responsibility for the accuracy and interpretation of the results, and the information presented in the text.\\

\noindent This research used data obtained with the Dark Energy Spectroscopic Instrument (DESI). DESI construction and operations is managed by the Lawrence Berkeley National Laboratory. This material is based upon work supported by the U.S. Department of Energy, Office of Science, Office of High-Energy Physics, under Contract No. DE–AC02–05CH11231, and by the National Energy Research Scientific Computing Center, a DOE Office of Science User Facility under the same contract. Additional support for DESI was provided by the U.S. National Science Foundation (NSF), Division of Astronomical Sciences, under Contract No. AST-0950945 to the NSF’s National Optical-Infrared Astronomy Research Laboratory; the Science and Technology Facilities Council of the United Kingdom; the Gordon and Betty Moore Foundation; the Heising-Simons Foundation; the French Alternative Energies and Atomic Energy Commission (CEA); the National Council of Humanities, Science and Technology of Mexico (CONAHCYT); the Ministry of Science and Innovation of Spain (MICINN), and by the DESI Member Institutions: \href{https://www.desi.lbl.gov/collaborating-institutions/}{www.desi.lbl.gov/collaborating-institutions}. The DESI collaboration is honored to be permitted to conduct scientific research on I’oligam Du’ag (Kitt Peak), a mountain with particular significance to the Tohono O’odham Nation. Any opinions, findings, and conclusions or recommendations expressed in this material are those of the authors and do not necessarily reflect the views of the U.S. National Science Foundation, the U.S. Department of Energy, or any of the listed funding agencies.\\

\noindent This work used the DiRAC@Durham facility managed by the Institute for Computational Cosmology on behalf of the STFC DiRAC HPC Facility (\href{www.dirac.ac.uk}{www.dirac.ac.uk}). The equipment was funded by BEIS capital funding via STFC capital grants ST/P002293/1, ST/R002371/1 and ST/S002502/1, Durham University and STFC operations grant ST/R000832/1. DiRAC is part of the National e-Infrastructure.

\pagebreak
\appendix

\section{Bispectrum window convolution}\label{app:convolution}

In this section, we outline the approximations and truncation criteria employed for the window-convolution series of the TripoSH bispectrum multipoles. We also assess the residual impact of fibre assignment after applying the window convolution in Section~\ref{ssec:convolution_test}.

\subsection{Convolution series}

The convergence of the window series depends on the specific footprint and scales. To determine which terms are the most relevant for the different tracers and bins we consider in this work, we adopt the weight function defined in \cite{wang2025} (their Equation~4.3),
\begin{equation}
    \gamma_{\ell_1 \ell_2 L}(\Lambda) = \frac{ \chi^2_{\ell_1 \ell_2 L}(\Lambda, \beta) - \chi^2_{\ell_1 \ell_2 L}(\Lambda_{\rm ref}, \beta) }{ \chi^2_{\ell_1 \ell_2 L}(\Lambda_0, \beta) -\chi^2_{\ell_1 \ell_2 L}(\Lambda_{\rm ref}, \beta) }.
    \label{eq:window_criteria}
\end{equation}

The weight function quantifies the difference between the $\chi^2$ of a convolution series $\Lambda$ that removes a single, specific contribution $Q_{\ell_1''\ell_2''L''}\zeta_{\ell_1'\ell_2'L'}$ from some reference case $\Lambda_{\rm ref}$. We stick to the choice presented in \cite{wang2025} for $\Lambda_{\rm ref}$:
\begin{align}
\hat{\zeta}_{000} & =
Q_{000} \zeta_{000}  \nonumber\\
& \hspace{0.5cm} + \frac{1}{3} Q_{110} \zeta_{110} + \frac{1}{5} Q_{220} \zeta_{220}  \nonumber\\
& \hspace{0.5cm} + \frac{1}{5} \left(Q_{022} \zeta_{022} + Q_{202} \zeta_{202}\right) + \frac{1}{6} Q_{112} \zeta_{112} + \frac{1}{9} \left(Q_{132} \zeta_{132} + Q_{312} \zeta_{312}\right)  \nonumber\\
& \hspace{0.5cm} - Q_{000} \bar{\zeta}, \\
\hat{\zeta}_{202} & =
Q_{000} \zeta_{202}  \nonumber\\
& \hspace{0.5cm} + Q_{202} \zeta_{000}  \nonumber\\
& \hspace{0.5cm} + \frac{1}{3} \left(Q_{112} \zeta_{110} + Q_{312} \zeta_{110} + Q_{110} \zeta_{112} + Q_{110} \zeta_{312}\right) + \frac{1}{5} \left(Q_{022} \zeta_{220} + Q_{220} \zeta_{022}\right)  \nonumber\\
& \hspace{0.5cm} + \frac{2}{7} Q_{202} \zeta_{202} + \frac{1}{6} \left(Q_{112} \zeta_{112} + Q_{132} \zeta_{132}\right) + \frac{8}{63} Q_{312} \zeta_{312} + \frac{1}{21} \left(Q_{312} \zeta_{112} + Q_{112} \zeta_{312}\right)  \nonumber\\
& \hspace{0.5cm} - Q_{202} \bar{\zeta}.
\end{align}

We follow \cite{wang2025}, and keep only the terms such that $|\gamma_{000}|, |\gamma_{202}| < 4 \times 10^{-4}$. These are summarised in Tab.~\ref{convolution_000_table}. As an example, for the SGC of the first LRG bin, the convolution series can be simplified as 
\begin{equation}
    \hat{\zeta}_{000} = Q_{\rm 000} \zeta_{\rm 000} + \frac{1}{3} Q_{\rm 110} \zeta_{\rm 110} + \frac{1}{5} Q_{\rm 022} \zeta_{\rm 022} + \frac{1}{5} Q_{\rm 202} \zeta_{\rm 202}
\end{equation}
and
\begin{align}
    \hat{\zeta}_{202} &= Q_{\rm 202} \zeta_{\rm 000} + \frac{1}{5} Q_{\rm 112} \zeta_{\rm 110} + \frac{1}{5} Q_{\rm 022} \zeta_{\rm 220} + \frac{1}{5} Q_{\rm 110} \zeta_{\rm 112} + \left(Q_{\rm 000} + \frac{2}{7} Q_{\rm 202}\right) \zeta_{\rm 202}.
\end{align}Given our metric, terms such as the $Q_{220}\zeta_{220}$ are irrelevant for the monopole convolution of this particular tracer and region of the sky.

\begin{table}
\centering
\caption{Coefficients in the convolution of the monopole $B_{000}$ for each tracer and sky region. The window coefficients $Q_{\ell_1'\ell_2'L'}$ multiply each 3PCF term $\zeta_{\ell_1^{''}\ell_2^{''}L^{''}}$ which are then summed in the window convolution series.}
\renewcommand{\arraystretch}{1.2}
\begin{tabular}{llccccc}
{}\\
\toprule
Sample & Region & $\zeta_{000}$ & $\zeta_{110}$ & $\zeta_{220}$ & $\zeta_{022}$ & $\zeta_{202}$ \\
\midrule
\multirow{2}{*}{LRG1} & SGC & $Q_{000}$ & $\tfrac13 Q_{110}$ & -- & $\tfrac15 Q_{022}$ & $\tfrac15 Q_{202}$ \\
& NGC & $Q_{000}$ & $\tfrac13 Q_{110}$ & $\tfrac15 Q_{220}$ & -- & -- \\
\midrule
\multirow{2}{*}{LRG2} & SGC & $Q_{000}$ & $\tfrac13 Q_{110}$ & $\tfrac15 Q_{220}$ & -- & -- \\
& NGC & $Q_{000}$ & $\tfrac13 Q_{110}$ & $\tfrac15 Q_{220}$ & -- & -- \\
\midrule
\multirow{2}{*}{LRG3} & SGC & $Q_{000}$ & $\tfrac13 Q_{110}$ & $\tfrac15 Q_{220}$ & -- & -- \\
& NGC & $Q_{000}$ & -- & -- & -- & -- \\
\midrule
\multirow{2}{*}{QSO} & SGC  & $Q_{000}$ & -- & $\tfrac15 Q_{220}$ & $\tfrac15 Q_{022}$ & $\tfrac15 Q_{202}$ \\
& NGC  & $Q_{000}$ & -- & -- & -- & -- \\
\bottomrule
\end{tabular}
\label{convolution_000_table}
\end{table}

\begin{table}
\centering
\caption{Same as Table \ref{convolution_000_table}, but for the quadrupole $B_{202}$. The integral-constraint contribution is omitted for brevity, as it is relevant only for the QSO sample, where it enters the convolution series as $-\zeta_{\rm ic} Q_{202}$.}
\renewcommand{\arraystretch}{1.2}
\begin{footnotesize}
\begin{tabular}{llccccccc}
{}\\
\toprule
Sample & Region & $\zeta_{000}$ & $\zeta_{110}$ & $\zeta_{220}$ & $\zeta_{112}$ & $\zeta_{022}$ & $\zeta_{202}$ & $\zeta_{312}$ \\
\midrule
\multirow{2}{*}{LRG1} & SGC & $Q_{202}$ & $\tfrac15 Q_{112}$ & $\tfrac15 Q_{022}$ & $\tfrac15 Q_{110}$ & -- & $Q_{000}+\tfrac27 Q_{202}$ & -- \\
& NGC & $Q_{202}$ & -- & -- & -- & -- & $Q_{000}$ & -- \\
\midrule
\multirow{2}{*}{LRG2} & SGC & $Q_{202}$ & $\tfrac13(Q_{112}+Q_{312})$ & $\tfrac15 Q_{022}$ & $\tfrac13 Q_{110}$ & -- & $Q_{000}+\tfrac27 Q_{202}$ & $\tfrac13 Q_{110}$ \\
& NGC & $Q_{202}$ & -- & -- & -- & -- & $Q_{000}+\tfrac27 Q_{202}$ & $\tfrac13 Q_{110}$ \\
\midrule
\multirow{2}{*}{LRG3} & SGC & $Q_{202}$ & -- & $\tfrac15 Q_{022}$ & -- & -- & $Q_{000}$ & -- \\
& NGC & $Q_{202}$ & -- & -- & $\tfrac13 Q_{110}$ & -- & $Q_{000}$ & -- \\
\midrule
\multirow{2}{*}{QSO} & SGC  & $Q_{202}$ & $\tfrac13(Q_{112}+Q_{312})$ & $\tfrac15 Q_{022}$ & $\tfrac13 Q_{110}$ & $\tfrac15 Q_{220}$ & $Q_{000}$ & $\tfrac13 Q_{110}$ \\
& NGC  & $Q_{202}$ & $\tfrac13(Q_{112}+Q_{312})$ & $\tfrac15 Q_{022}$ & $\tfrac13 Q_{110}$ & $\tfrac15 Q_{220}$ & $Q_{000}+\tfrac27 Q_{202}$ & $\tfrac13 Q_{110}$ \\
\bottomrule
\end{tabular}
\end{footnotesize}
\end{table}

\paragraph{Window matrix} In practice, the window-convolution operator ${\sf W}_B = {\sf T}_2\,{\sf C}\,{\sf T}_1$, introduced in Section~\ref{ssec:window_convolution}, is constructed with \texttt{Triumvirate}. The operator ${\sf W}_B$ is evaluated on the canonical basis ${\bs e}_\lambda \otimes \bs{e}_{ij}$ of the finite-dimensional vector space spanned by the input bispectrum multipoles $\lambda\equiv(\ell_1,\ell_2,L)$ and the $(k_i,k_j)$ bins. The resulting convolved vectors define the corresponding window-convolution matrix ${\sf W}_B$ in this basis, such that any theory vector can then be convolved as $\hat{\bs{B}}={\sf W}_B\,\bs{B}$ (see \cite{wang2025} for more details). 

\begin{figure}[htpb]
    \centering
    \includegraphics[width=\linewidth]{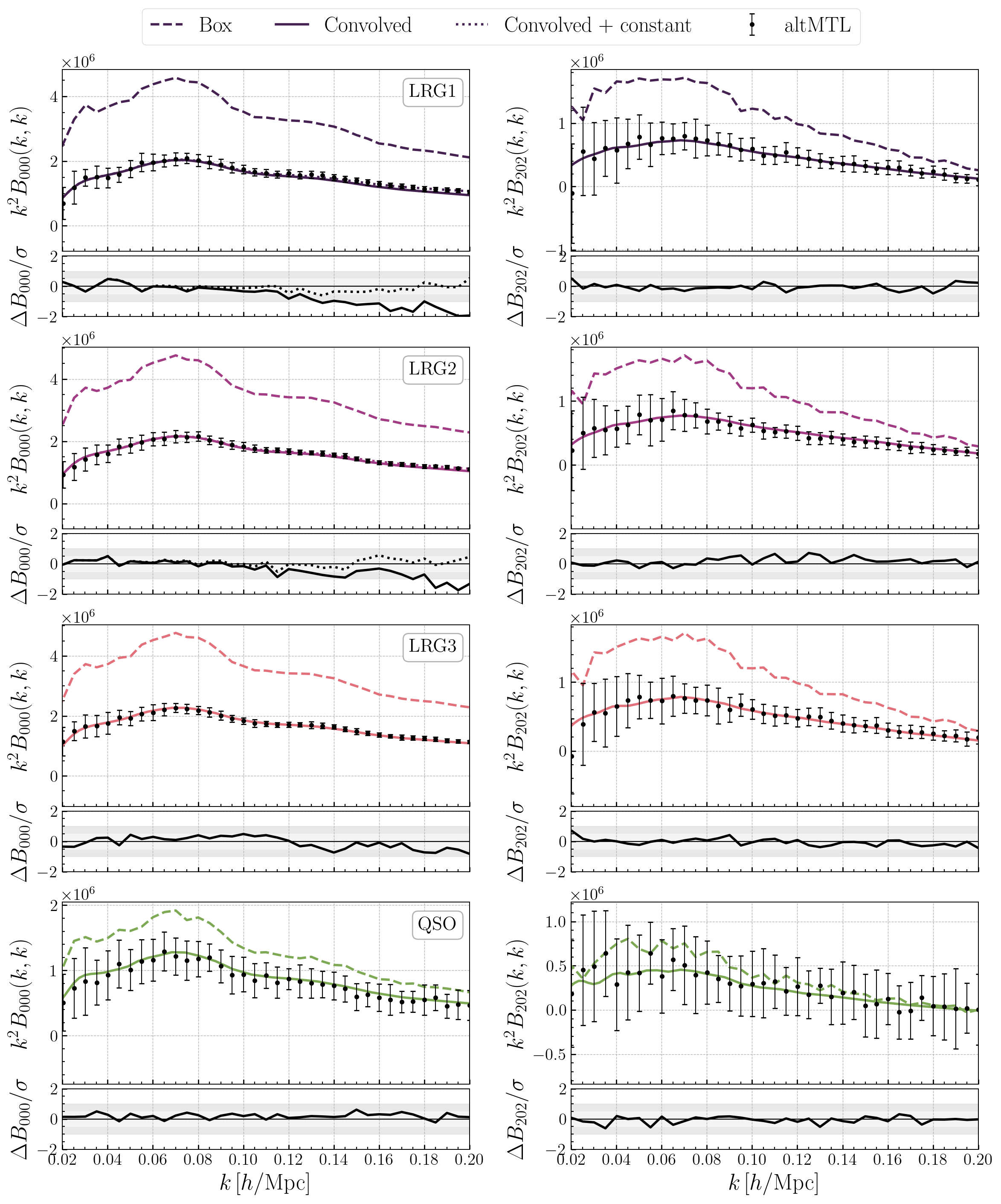}
    \vspace{-8mm}
    \caption{Validation of the bispectrum window convolution using the mean of 25 altMTL mocks (black points; errorbars show the standard deviation of the measurements). The left (right) panels show the diagonal bispectrum monopole (quadrupole) for the GC-combined bins. Dashed curves show the mean bispectrum measured from the corresponding 25 boxes, used as a proxy for the theory. Bottom panels show the residuals between the measurements and the convolved prediction, normalised by the standard deviation of the 25 altMTL mocks, with (solid) and without (dotted) the constant offset included for the LRG1 and LRG2 bins.~Shaded bands indicate the $\pm 0.5\sigma$ (light) and $\pm 1\sigma$ (dark) regions.~Bispectrum units are omitted for clarity.}
    \label{fig:convolution_test}
\end{figure}

\subsection{Convolution test}\label{ssec:convolution_test}

We validate the bispectrum window convolution using the mean of the 25 altMTL mocks (see Section~\ref{ssec:nbody}), for all tracers considered in this analysis, as shown in Figure~\ref{fig:convolution_test}. To assess the accuracy of the window convolution, the theory vector $\bs{B}$ is obtained from the mean of the 25 boxes from which the cut-sky mocks were generated, as they come from the same realisations. 

For this test, we considered the GC-combined data vectors and window matrices, as described in Section~\ref{sec:data}. These are shown in Figure~\ref{fig:convolution_test} as the black dots. The error bars were obtained from the standard deviation $\sigma$ of the 25 measurements. We can see that both bispectrum multipoles are accurately reproduced through the window convolution procedure over the whole range of scales we consider in this work ($k_{\rm max} = 0.2\,\,h/{\rm Mpc}$), as can be seen in the residuals $\Delta B_{\ell_1 \ell_2 L}/\sigma$.

For the first two LRG bins, we found the remaining monopole mismatch to be mostly scale independent. In order to check for that, we have added a single constant offset to the convolution of the cubic-box measurements of these bins. The results are within the 1-$\sigma$ region, indicating that this shot-noise-like contribution can be used to bypass residuals of the window convolution. The constant $P_{\rm shot}$ parameter (Equation~\ref{eq:bispectrum_model}), varied in our fits, should absorb the amplitude offset at scales $k < 0.12\,\,h/{\rm Mpc}$ for these two redshift bins. In Appendix~\ref{app:validation}, however, we show for the most extreme case (LRG1) that the posterior constraints are robust up to $k_{\rm max} = 0.2\,\,{h/{\rm Mpc}}$.

\begin{figure}[ht]
    \centering
    \includegraphics[width=\linewidth]{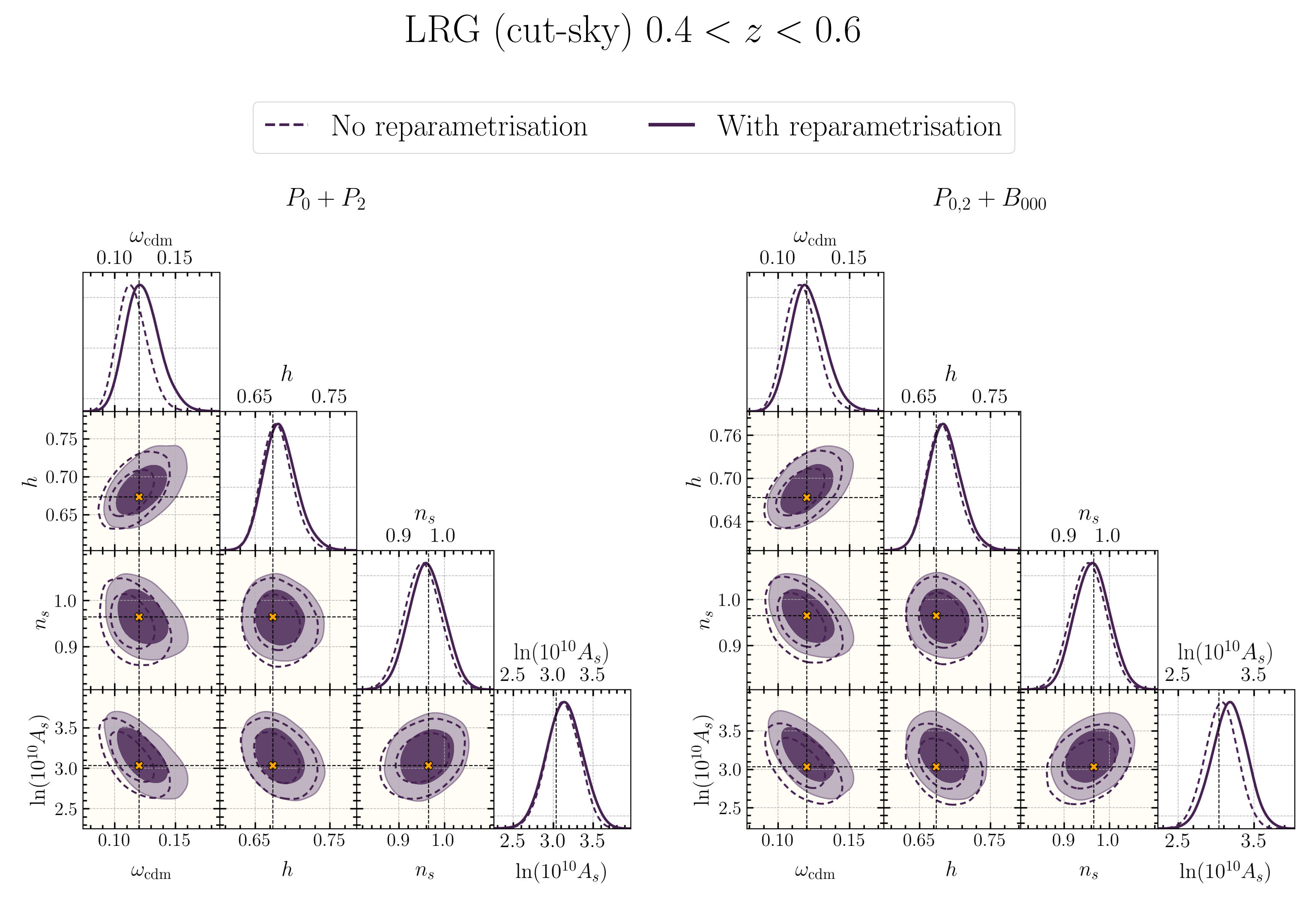}
    \caption{Constraints on the cosmological parameters obtained from the mean of 25 cut-sky \textsc{AbacusSummit} altMTL mock realisations, in the redshift range $0.4<z<0.6$. The left panel shows the constraints derived by considering the power-spectrum monopole $P_0$ and quadrupole $P_2$ only, whilst the right panel additionally includes the bispectrum monopole $B_{000}$ in the data vector. Solid and dashed contours correspond to analyses performed with and without the nuisance-parameter reparametrisation described in Section~\ref{sssec:priors}, respectively. The crosses indicate the fiducial cosmology of the mocks.}
    \label{fig:LRG1_mock_test}
\end{figure}

\section{Validation on mocks}\label{app:validation}
In this Appendix, we summarise the main validation tests carried out on the mean of the 25 cut-sky \textsc{AbacusSummit} mocks.

\subsection{Impact of reparametrisation}

In Figure~\ref{fig:LRG1_mock_test} we show the impact of the reparametrisation adopted in this work, using the mean of the 25 LRG1 cut-sky mocks. As discussed in Section~\ref{sssec:priors}, the reparametrisation is applied to reduce prior-volume effects arising from cosmology-dependent amplitudes in the biases and EFT parameters. 

We used the LRG1 sample as it corresponds to the largest differences in $\omega_{\rm cdm}$ with respect to the DESI 2024 power-spectrum-only constraints. As already pointed out in \cite{tsedrik2026}, prior-volume effects favour lower $A_{\rm AP}$. Hence, by removing such effects we would expect larger $A_{\rm AP}$ which is directly related to a higher $\omega_{\rm cdm}$, for a well-constrained Hubble factor $h$. This is observed in the posteriors shown in Figure~\ref{fig:LRG1_mock_test}. 

\subsection{Choice of maximum scale}

To test the dependence of the cosmological constraints on the maximum bispectrum monopole scale $k_{\rm max}^{000}$, we have fixed the power-spectrum monopole and quadrupole cut-off to $k_{\rm max}^{(0,2)}=0.20\,\,h/{\rm Mpc}$ and varied $k_{\rm max}^{000}$ between $0.14$ and $0.20\,\,h/{\rm Mpc}$. All cosmological parameters remained within approximately 1.3$\sigma$ of their fiducial values, and their variation across the different scale cuts was smaller than 0.4$\sigma$. The largest displacement occurred for $\ln(10^{10}A_s)$ in LRG3 and QSO, with an average of $\sim 1.2\sigma$ shift towards higher values compared to the fiducial $\ln(10^{10}A_s) = 3.036$. However, it did not increase as smaller scales were included. We therefore found no evidence for a scale-dependent parameter shift over the range considered.

We have found that the information gained from extending the bispectrum analysis to $k_{\rm max} = 0.20\,\,h/{\rm Mpc}$ is concentrated primarily in $\omega_{\rm cdm}$, whose uncertainty decreased by approximately 12\%, 12\%, 8\% and 7\% for LRG1, LRG2, LRG3, and QSO respectively, relative to $k_{\max}^{000}=0.14\,\,h/{\rm Mpc}$. Changes in the remaining cosmological uncertainties are generally below 7\% and are not consistently strengthened (or weakened) by increasing (or decreasing) the cut-off scale. These tests therefore show no evidence of a scale-dependent parameter bias and support the choice of $k_{\rm max}^{000}=0.20\,\,h/{\rm Mpc}$, while also showing that most of the additional constraining power is associated with the determination of $\omega_{\rm cdm}$.

\begin{figure}[ht]
    \centering
    \includegraphics[width=\linewidth]{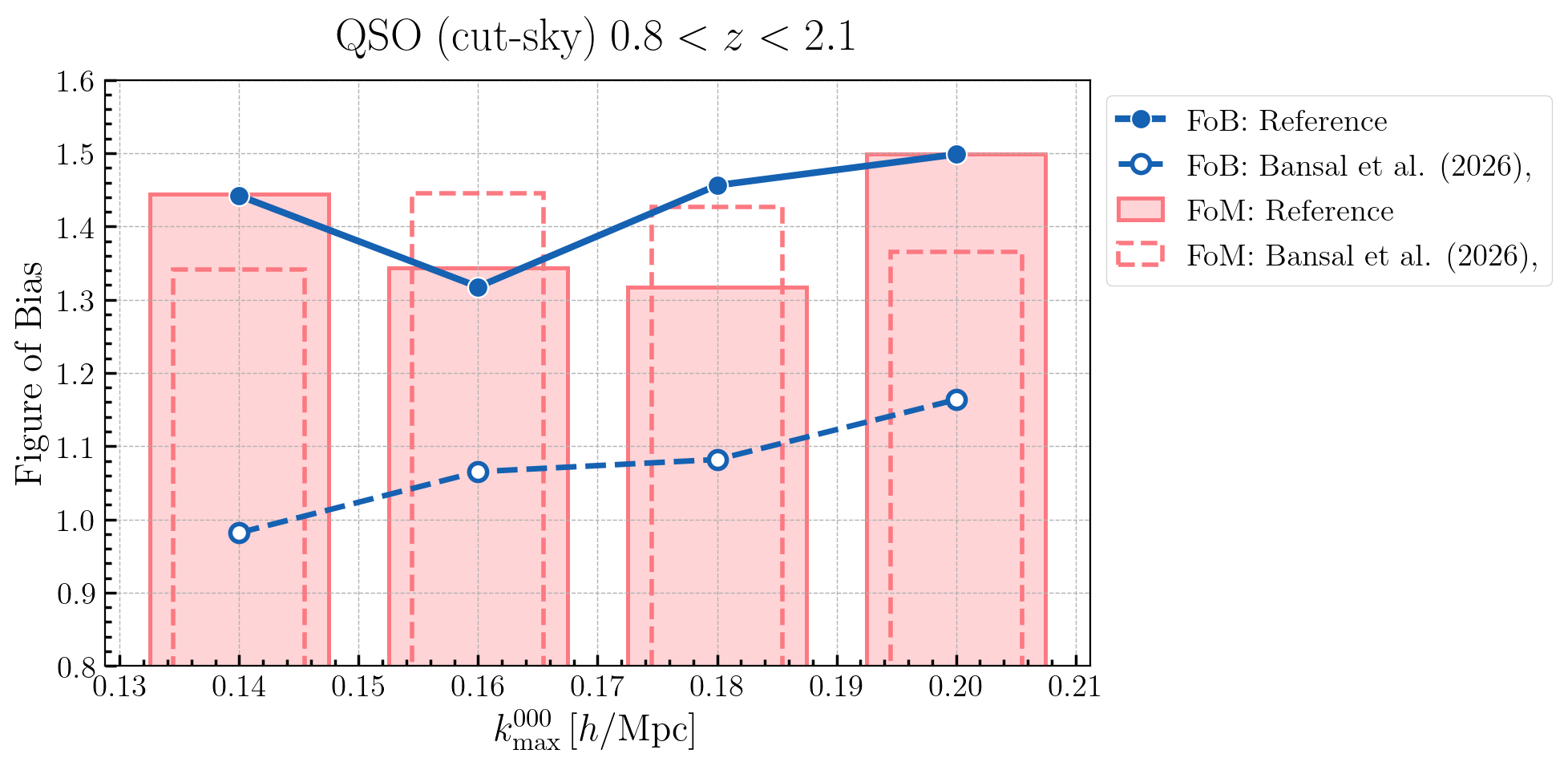}
    \caption{Figure of Bias (FoB) and Figure of Merit (FoM) as a function of the bispectrum monopole maximum scale $k_{\rm max}^{000}$ for the QSO cut-sky mocks. The power spectrum monopole and quadrupole $k_{\rm max}^{(0,2)} = 0.2\,\,h/{\rm Mpc}$ were fixed. The solid blue line shows the FoB for the reference priors of Table~\ref{tab:priors}, whilst the dashed line shows the same analysis considering the priors used in Bansal et al. (2026) \cite{bansal2026}. The same applies to the rescaled FoM.}
    \label{fig:FoBFoM}
\end{figure}

Finally, we have computed the figure of bias (FoB) and figure of merit (FoM), defined as \citep{eggemeier2020}
\begin{equation}
    {\rm FoM} \equiv \left({\rm det}\,\tilde{\sf C}_{\theta_\alpha,\theta_\beta}\right)^{-1/2},\text{ and } {\rm FoB} \equiv \left[\sum_{\alpha,\beta}(\bar{\theta}_\alpha-\theta_\alpha^{\rm fid})\rm{\sf C}^{-1}_{\theta_\alpha,\theta_\beta}\,(\bar{\theta}_\beta-\theta_\beta^{\rm fid})\right]^{1/2},
\end{equation}where $\tilde{\sf C}_{\theta_\alpha,\theta_\beta} \equiv {\sf C}_{\theta_\alpha,\theta_\beta}/\theta_\alpha^{\rm fid}\,\theta_\beta^{\rm fid}$ is the normalised covariance of the parameters $\theta_\alpha$, for the priors shown in Table~\ref{tab:priors}. For FolpsD-inspired priors, the priors on the power-spectrum counterterms are tighter.

\clearpage

\section{Extra figures}\label{app:extra}
For completeness, in this Appendix we gather the figures showing the posterior distributions obtained from the single-tracer analyses. The 1-$\sigma$ constraints are quoted in Table~\ref{tab:main_results}.

\begin{figure}[H]
    \centering
    \includegraphics[width=\linewidth]{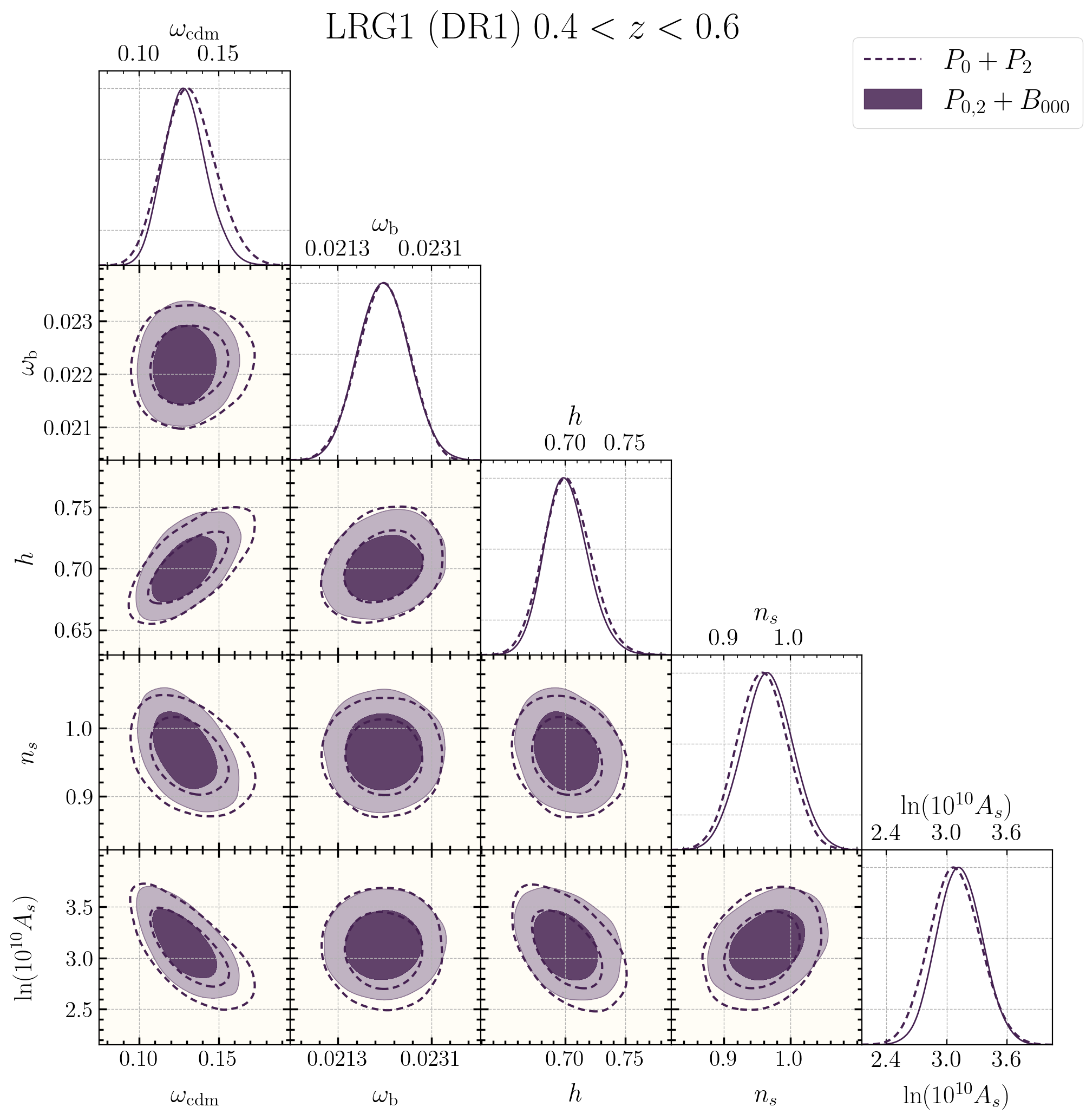}
    \caption{Posterior constraints obtained from the DR1 LRG1 bin ($0.4<z<0.6$). Dashed contours show the constraints obtained from the power-spectrum-only analysis, while the shaded ones include the bispectrum monopole.}
    \label{fig:LRG1_results}
\end{figure}

\begin{figure}[H]
    \centering
    \includegraphics[width=\linewidth]{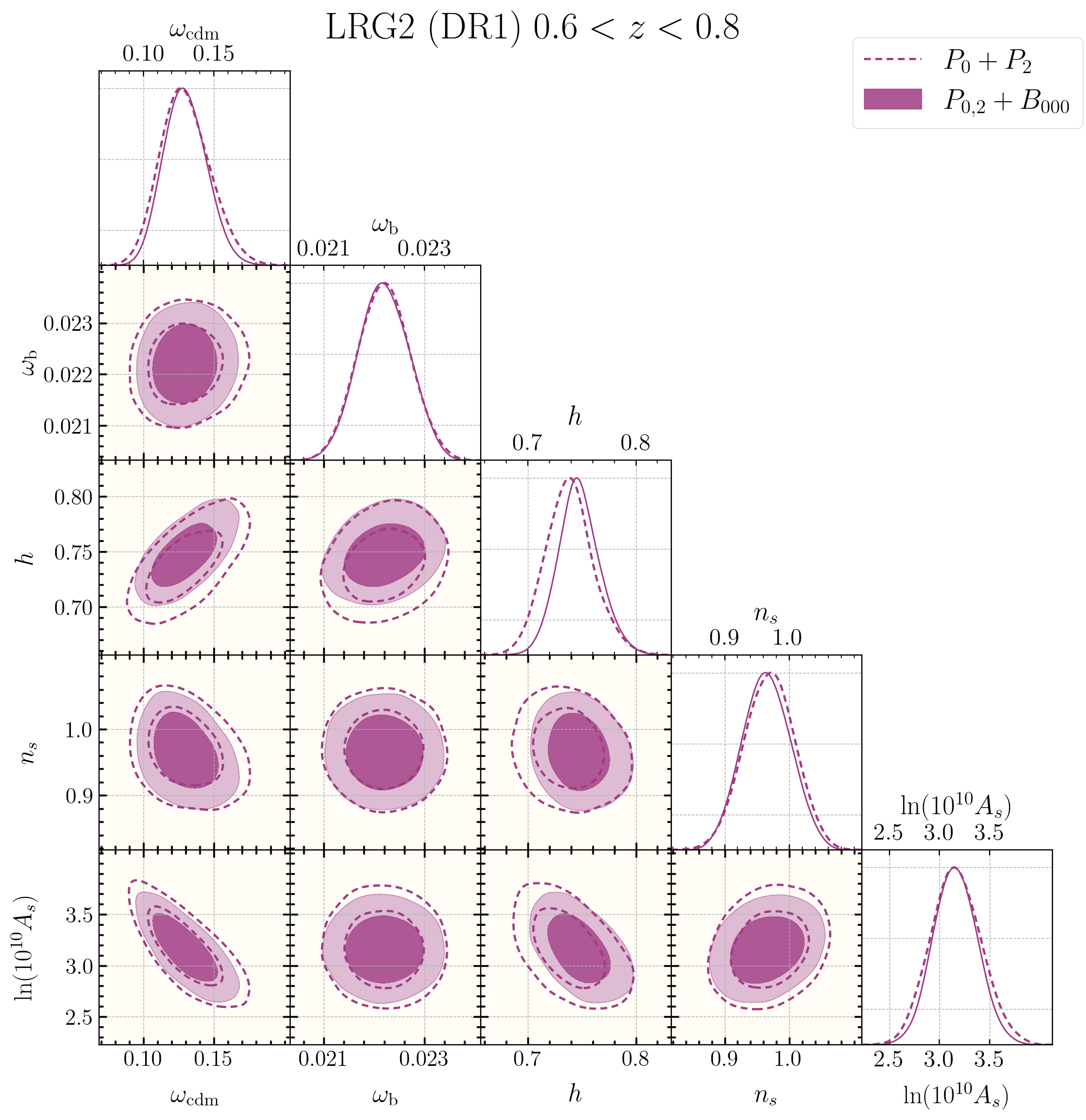}
    \caption{Posterior constraints obtained from the DR1 LRG2 bin ($0.6<z<0.8$). Dashed contours show the constraints obtained from the power-spectrum-only analysis, while the shaded ones include the bispectrum monopole. }
    \label{fig:LRG2_results}
\end{figure}

\begin{figure}[H]
    \centering
    \includegraphics[width=\linewidth]{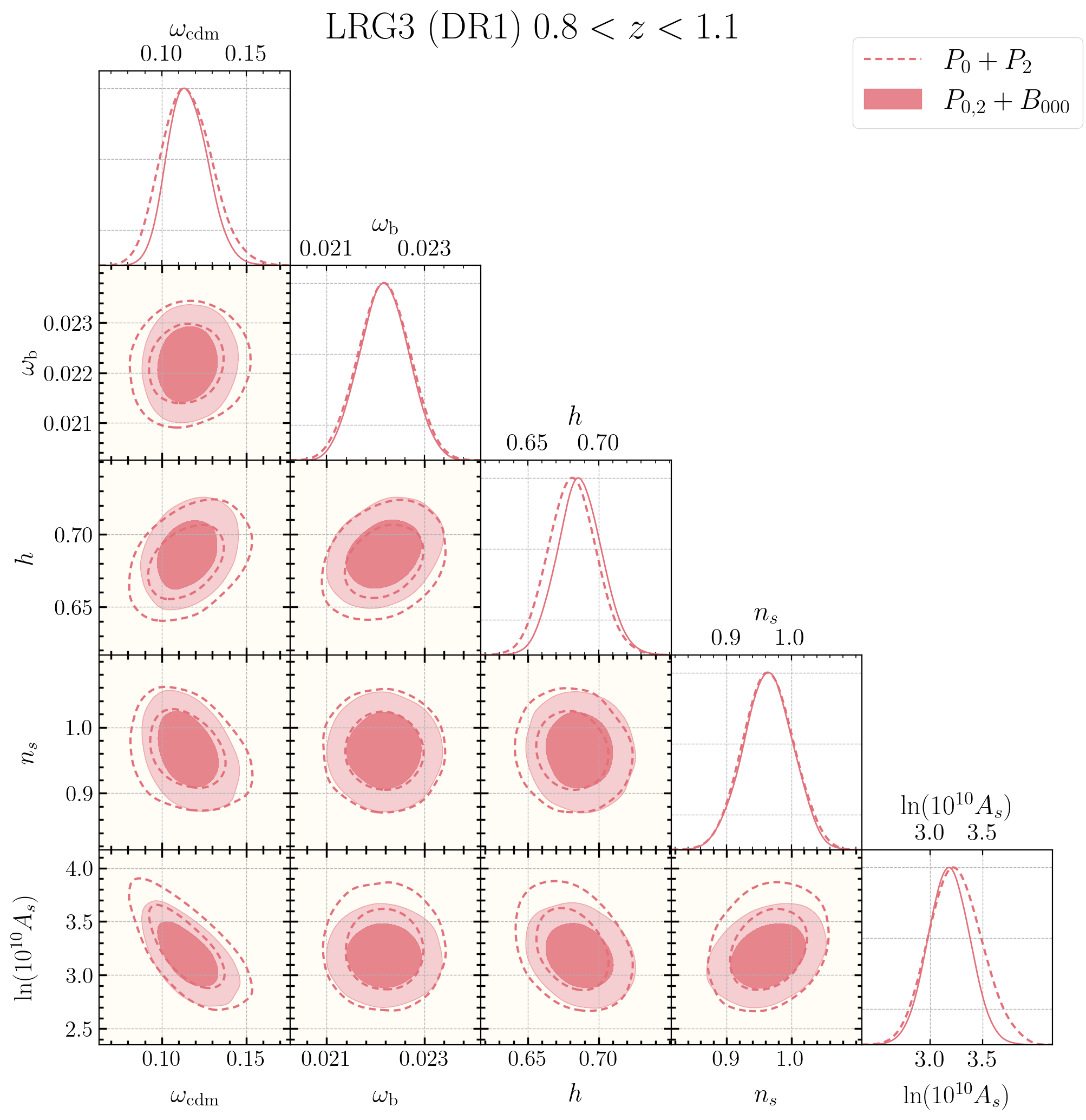}
    \caption{Posterior constraints obtained from the DR1 LRG3 bin ($0.8<z<1.1$). Dashed contours show the constraints obtained from the power-spectrum-only analysis, while the shaded ones include the bispectrum monopole.}
    \label{fig:LRG3_results}
\end{figure}

\begin{figure}[H]
    \centering
    \includegraphics[width=\linewidth]{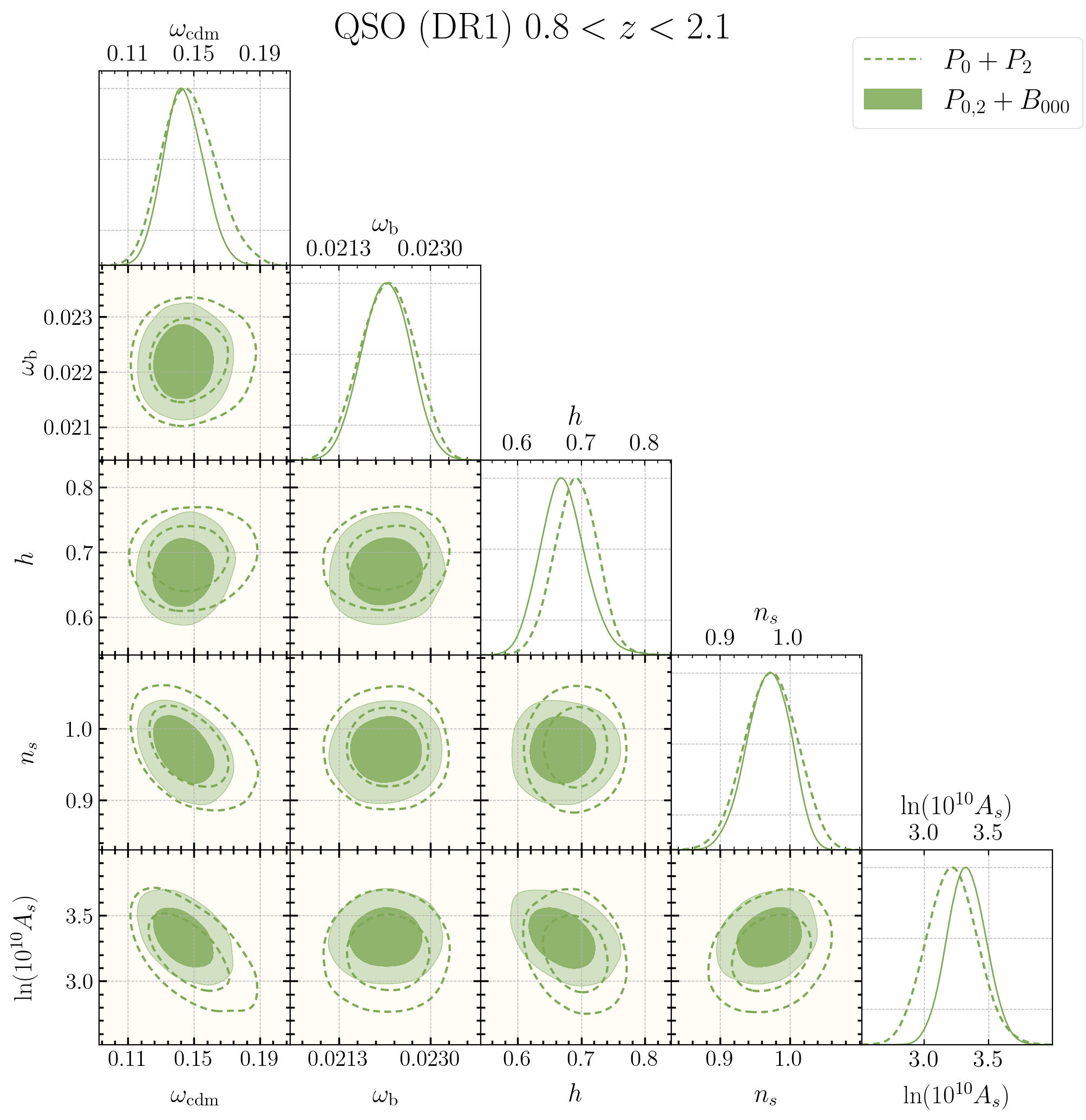}
    \caption{Posterior constraints obtained from the DR1 QSO bin ($0.8<z<2.1$). Dashed contours show the constraints obtained from the power-spectrum-only analysis, while the shaded ones include the bispectrum monopole.}
    \label{fig:QSO_results}
\end{figure}

\pagebreak

\bibliographystyle{JHEP}
\bibliography{main}

\end{document}